\documentclass[letterpaper,twocolumn,10pt]{article}
\usepackage{zhanggroup}

\usepackage{url}
\usepackage{tikz}
\usepackage{amsfonts}
\usepackage{xspace} 
\usepackage{amsmath}
\usepackage{mathrsfs}
\usepackage{amsthm}
\usepackage{wrapfig}
\usepackage{xifthen}
\usepackage{booktabs}
\usepackage{multirow}
\usepackage[hang,flushmargin]{footmisc}
\usepackage{makecell}
\usepackage[absolute]{textpos}
\usepackage{colortbl}
\usepackage{etoolbox}
\usepackage{hhline}
\usepackage{algorithm}
\usepackage{algorithmic}
\usepackage{adjustbox}
\usepackage[most]{tcolorbox}
\usepackage[font=scriptsize, labelfont=bf, belowskip=-7pt]{subcaption}
\newcommand{\mypara}[1]{\noindent{\bf {#1}.}}
\newcommand{\myrq}[1]{\noindent{\bf {#1}:}}
\newcommand{\systemname}{\textsc{CoAT}\xspace}

\newcommand{\adv}{\ensuremath{\mathsf{ADV}}\xspace}
\newcommand{\atoa}[2]{\ensuremath{{#1}\allowbreak2\allowbreak{#2}}\xspace}
\newcommand{\atoatoa}[3]{\ensuremath{{#1}\allowbreak2\allowbreak{#2}\allowbreak2\allowbreak{#3}}\xspace}
\newcommand{\meminf}{\ensuremath{\mathsf{MemInf}}\xspace}
\newcommand{\propinf}{\ensuremath{\mathsf{PropInf}}\xspace}
\newcommand{\attrinf}{\ensuremath{\mathsf{AttrInf}}\xspace}
\newcommand{\aux}{\mathsf{aux}}
\newcommand{\target}{\mathsf{target}}
\newcommand{\xtarget}{\ensuremath{x_{\target}}}
\newcommand{\xattack}[1]{\ensuremath{x_{\mathsf{{#1}}}}}
\newcommand{\BM}{\ensuremath{\mathcal{M}^{\mathsf{B}}}\xspace}
\newcommand{\WM}{\ensuremath{\mathcal{M}^{\mathsf{W}}}\xspace}
\newcommand{\LIRA}{\ensuremath{\mathsf{LiRA}}\xspace}
\newcommand{\QD}{\ensuremath{\mathcal{D}_{\aux}^{\mathsf{Q}}}\xspace}
\newcommand{\SD}{\ensuremath{\mathcal{D}_{\aux}^{\mathsf{S}}}\xspace}
\newcommand{\PD}{\ensuremath{\mathcal{D}_{\aux}^{\mathsf{P}}}\xspace}
\newcommand{\model}[2]{\ensuremath{\mathcal{M}_{#1}^{#2}}\xspace}
\newcommand{\dset}[2]{\ensuremath{\mathcal{D}_{#1}^{#2}}\xspace}
\newcommand{\atktu}[3]{\ensuremath{\langle {#1},{#2},{#3} \rangle}\xspace}
\newcommand{\tuple}[1]{\ensuremath{\langle #1 \rangle}}

\newcommand{\customTableFont}{\fontsize{9pt}{9.5pt}\selectfont}

\begin{document}
%-------------------------------------------------------------------------------

\date{}

%-------------------------------------------------------------------------------
\title{\Large \bf Amplifying Machine Learning Attacks Through Strategic Compositions}
%-------------------------------------------------------------------------------

\author{
{\rm Yugeng Liu\textsuperscript{1}}\ \ \
{\rm Zheng Li\textsuperscript{2}}\ \ \
{\rm Hai Huang\textsuperscript{1}}\ \ \
{\rm Michael Backes\textsuperscript{1}}\ \ \
{\rm Yang Zhang\textsuperscript{1}}\ \ \
\\
\\
\textsuperscript{1}\textit{CISPA Helmholtz Center for Information Security}\ \ \ 
\textsuperscript{2}\textit{Shandong University}\ \ \
}

\maketitle
%-------------------------------------------------------------------------------

%-------------------------------------------------------------------------------
\begin{abstract}
%-------------------------------------------------------------------------------

Machine learning (ML) models are proving to be vulnerable to a variety of attacks that allow the adversary to learn sensitive information, cause mispredictions, and more. 
While these attacks have been extensively studied, current research predominantly focuses on analyzing each attack type individually.
In practice, however, adversaries may employ multiple attack strategies simultaneously rather than relying on a single approach.
This prompts a crucial yet underexplored question: When the adversary has multiple attacks at their disposal, are they able to mount or amplify the effect of one attack with another?
In this paper, we take the first step in studying the strategic interactions among different attacks, which we define as \emph{attack compositions}. 
Specifically, we focus on four well-studied attacks during the model's inference phase: adversarial examples, attribute inference, membership inference, and property inference. 
To facilitate the study of their interactions, we propose a taxonomy based on three stages of the attack pipeline: preparation, execution, and evaluation.
Using this taxonomy, we identify four effective attack compositions, such as property inference assisting attribute inference at its preparation level and adversarial examples assisting property inference at its execution level. 
We conduct extensive experiments on the attack compositions using three ML model architectures and three benchmark image datasets.
Empirical results demonstrate the effectiveness of these four attack compositions.
We implement and release a modular reusable toolkit, \systemname.
Arguably, our work serves as a call for researchers and practitioners to consider advanced adversarial settings involving multiple attack strategies, aiming to strengthen the security and robustness of AI systems.

%-------------------------------------------------------------------------------
\end{abstract}
%-------------------------------------------------------------------------------

%-------------------------------------------------------------------------------
\section{Introduction}
%-------------------------------------------------------------------------------

Recently, machine learning has gained momentum in multiple fields, achieving success in real-world deployments, such as image classification~\cite{DCLT19,BHWWW20,ZLCYJL21}, face recognition~\cite{ZDH17,KSMB16}, and medical image analysis~\cite{KEEKF15,SWFJH10,BFDB11}.
Nevertheless, prior research has shed light on the vulnerability of ML models to various attacks, such as adversarial examples~\cite{IWGZ18,RSG18,ASEHSC18}, membership inference~\cite{SSSS17,NSH18,SZHBFB19,LZ21}, and backdoor attacks~\cite{CLLLS17,GDG17,LMALZWZ18}. 
These vulnerabilities prompt significant security and privacy risks.
As a result, investigating, quantifying, and mitigating these various attacks on ML models have become increasingly important topics.

\begin{figure}[!t]
\centering
\includegraphics[width=0.95\columnwidth]{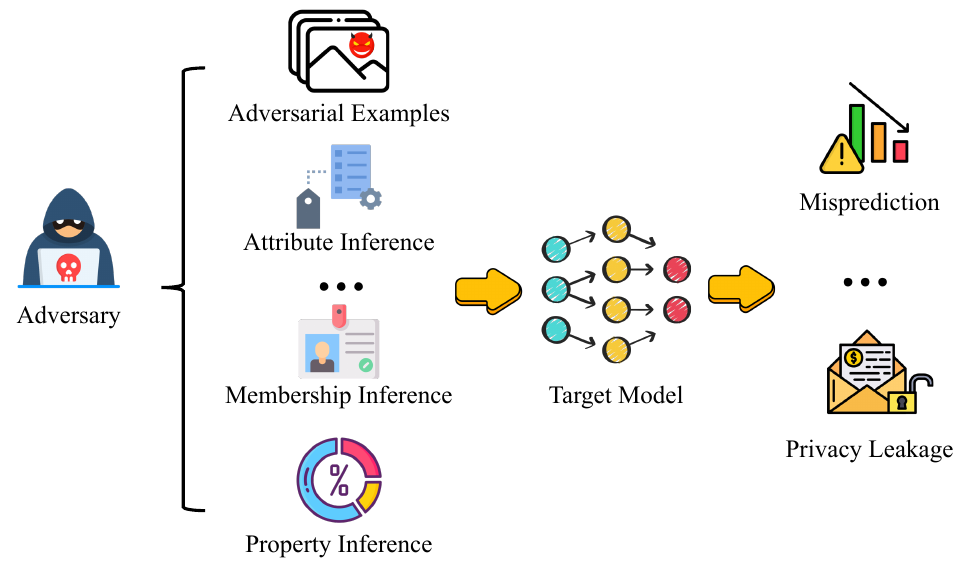}
\caption{Given a target model, the adversary can launch different attacks to achieve different malicious goals.}
\label{figure:attacks}
\end{figure}

Currently, most research in this field focuses on developing or optimizing more powerful attacks, e.g., higher attack success rates or greater stealthiness, and proposing corresponding countermeasures.
More precisely, these studies typically focus on individual attacks.
While some measurement or benchmark papers exist that consider multiple attacks, e.g., ML-Doctor~\cite{LWHSZBCFZ22} or SecurityNet~\cite{ZLYHBFZ24}, they still implement each attack individually.
In other words, studying attacks in isolation is actually the most common practice in the existing ML security domain.

However, this practice may not accurately reflect real-world scenarios, where adversaries often possess multiple attack strategies and can potentially synergize or leverage them simultaneously. 
When focusing solely on individual attacks, researchers may overlook the potential for adversaries to amplify the impact of one attack by leveraging knowledge or capabilities gained from another attack. 
Consequently, the true extent of vulnerabilities and risks posed by combined attacks may be underestimated or remain unexplored.
This reality prompts the need for a more comprehensive understanding of the \textit{intentional interactions} among different attacks.

\subsection{Contributions}
In this work, we take the first step in exploring the (possible) intentional interactions between different types of attacks.
We focus exclusively on the inference phase of ML models since deployed models are more likely to face intentional interactions between different attacks.
Specifically, we consider the four most representative attacks launched during the inference phase of the ML model, aka \textit{inference time attacks}: adversarial examples~\cite{IWGZ18,RSG18,ASEHSC18}, attribute inference~\cite{MSCS19,SS20}, membership inference~\cite{SSSS17,NSH18,SZHBFB19,LZ21}, and property inference~\cite{MSCS19}.

We formulate the following research questions (RQs), aiming at addressing this significant gap.

\begin{itemize}
\item \myrq{RQ1} How can we approach the design and implementation of attack compositions?
\item \myrq{RQ2} How can the knowledge gained from one type of attack facilitate or amplify the effectiveness of another attack?
\item \myrq{RQ3} How effective are combined attacks in exploiting ML model vulnerabilities compared to individual ones?
\end{itemize} 

\mypara{Composition Taxonomy} 
First, we propose a taxonomy for attack compositions based on the attack pipeline (\textbf{RQ1}), divided into three levels: preparation, execution and evaluation.
The former encompasses all preliminary activities before the main attack, including tool setup, data collection, and configuration. 
The execution level covers the attack's actual implementation, involving malicious queries, responses, and vulnerability exploitation.
Finally, the evaluation level assesses the attack impact, including system disruption, goal achievement, and any post-exploitation activities.

\mypara{Composition Methodology}
Based on the taxonomy, we conduct an extensive exploration of attack compositions across four representative inference-time attacks (\textbf{RQ2}). 
Specifically, we identify four effective attack compositions: one at the preparation level, two at the execution level, and one at the assessment level.
At the preparation level, we propose using property inference to assist attribute inference (\atoa{\propinf}{\attrinf}). 
By determining the attribute distribution in the victim model's training dataset through property inference, we use it to create a balanced attack training dataset for attribute inference.
At the execution level, we propose two attack compositions: using adversarial examples to assist membership inference (\atoa{\adv}{\meminf}) and property inference (\atoa{\adv}{\propinf}), respectively. Adversarial examples can search for different noise magnitudes for various membership or property statuses, which are then integrated into their original information for improved attack performance.
At the evaluation level, we leverage property inference to assist membership inference (\atoa{\propinf}{\meminf}). After the membership inference process ends, we use the property distribution determined by property inference to calibrate its attack output.

\mypara{Composition Evaluation} We conduct extensive experiments across three popular ML model architectures and three benchmark image datasets (\textbf{RQ3}).
Here, we summarize our analysis using ResNet18~\cite{HZRS16} trained on CIFAR10~\cite{CIFAR} as an example.
First, property inference significantly amplifies attribute inference at its preparation level.
For instance, \attrinf achieves an accuracy of 0.500, while \atoa{\propinf}{\attrinf} achieves an empirical accuracy of 0.894 and a theoretical accuracy of 0.872.
Second, adversarial examples improve both membership inference and property inference.
For instance, the black-box \meminf with shadow model and \propinf achieve an accuracy of 0.664 and 0.890, respectively, while the attack compositions yield significantly improved results, with accuracies of 0.851 and 0.960, respectively.
Finally, the black-box \meminf with partial training dataset achieves an accuracy of 0.631, compared to \atoa{\propinf}{\meminf}'s accuracy of 0.669.

\mypara{\systemname}
To evaluate our proposed diverse attack compositions, we develop a modular framework, \systemname (\underline{Co}mposition of \underline{AT}tacks).
With its modular design, \systemname allows for easy integration of new versions of each attack type, additional datasets, and models. 
Our code will be released publicly along with the final version of the paper (and is already available upon request), thus facilitating further research in the field. 

\mypara{Note}
We deliberately exclude model-stealing attacks from consideration, as the process of stealing essentially transforms a black-box model into a white-box model, which falls outside our formal definition of attack composition. 
Additionally, we omit consideration of training-time attacks, such as backdoor attacks, since these scenarios presuppose adversarial involvement in the training process of victim models -- a condition that violates our fundamental assumptions about attacker capabilities, definition of compositions, and access limitations.

%-------------------------------------------------------------------------------
\section{Inference-Time Attacks}
\label{section:preliminary}
%-------------------------------------------------------------------------------

In this section, we present the four most representative attacks during the ML models' inference phase, namely, adversarial examples (\autoref{section:adv}), attribute inference (\autoref{section:attrinf}), membership inference (\autoref{section:meminf}), and property inference (\autoref{section:propinf}).
Specifically, the first three are designed at the sample level, while the last one aims to infer the general information at the dataset level.
Different attacks can be applied to different threat models; see \autoref{table:taxonomy}.
For each attack and each threat model, we focus on one representative state-of-the-art method.

\begin{table}[h]
\centering
\customTableFont
\setlength{\tabcolsep}{10 pt}
\caption{Different attacks under different threat models.}
\label{table:taxonomy}
\begin{tabular}{@{}l@{}l | c c@{}}
\toprule
\multicolumn{2}{c|}{\bf Auxiliary} & \multicolumn{2}{c}{\bf Model Access}\\
\multicolumn{2}{c|}{\bf Dataset} & Black-Box (\BM) & White-Box (\WM) \\
\midrule
Partial & (\PD) & \meminf & \meminf, \attrinf \\
Shadow~ &(\SD) & \meminf, \propinf & \meminf, \attrinf \\
Query & (\QD) & \propinf & - \\
\bottomrule
\end{tabular}
\end{table}

%-------------------------------------------------------------------------------
\subsection{Adversarial Examples}
\label{section:adv}
%-------------------------------------------------------------------------------

Adversarial examples (\adv)~\cite{SZSBEGF14, GSS15,CW17,GSS15,PMJFCS16,MMSTV18,IWGZ18,RSG18,ASEHSC18,BB18} are a type of ML security threat where malicious inputs are deliberately designed to deceive ML models. 
These inputs, known as adversarial examples, are typically crafted by making small, often imperceptible modifications to target data to cause the model to predict incorrectly. 
More formally, given a target data sample $\xtarget$, (the access to) a target model $\model{}{}$, an adversarial example $\xattack{adv}$ can be generated by applying a perturbation $\delta$ such that $\xattack{adv} = \xtarget + \delta$.
To ensure it remains subtle, the perturbation is usually constrained by a norm $\| \delta \|_p \leq \epsilon$.
The goal is to maximize the loss function $\ell\left(\model{\theta}{}(\xattack{adv}), y\right)$
In general, an adversarial attack can be defined as:
\begin{equation}
\begin{split}
    \adv:\xtarget,\model{}{}\rightarrow \{\xattack{adv} \}
\end{split}
\end{equation}
In general, this type of attack can be categorized into two types based on the knowledge of the adversary: black-box and white-box attacks ($\model{}{}\in\{\BM,\WM\}$).

\mypara{Black-Box \atktu{\adv}{\BM}{\xtarget}~\cite{ACFH20}}
Black-box attacks operate under the assumption that the adversary has no internal knowledge of the models. 
Instead, the adversary can only observe the outputs from the model. 
This scenario is more common in the real world, where internal details are inaccessible.
They usually leverage trial-and-error to approximate the gradient of the target model~\cite{CJW20} or randomized search schemes to approximate the boundary of the data samples~\cite{ACFH20}.

\mypara{White-Box \atktu{\adv}{\WM}{\xtarget}~\cite{MMSTV18}}
White-box attacks assume the adversary has complete knowledge of the model, including its architecture, parameters, and training data.
It allows the adversary to precisely calculate the most effective perturbations to maximize errors of ML models, often employing gradient-based methods to manipulate the input data directly, such as C\&W~\cite{CW17}, FGSM~\cite{GSS15}, JSMA~\cite{PMJFCS16}, and PGD~\cite{MMSTV18}.

%-------------------------------------------------------------------------------
\subsection{Attribute Inference}
\label{section:attrinf}
%-------------------------------------------------------------------------------

During the training phase, an ML model might unintentionally learn information that is not directly relevant to its intended tasks.
If these models are open-source and published on the internet, the adversary may use them to predict some sensitive information.
For example, a model designed to predict some features such as eye color from profile pictures could inadvertently also develop the ability to leak ethnicities~\cite{MSCS19,SS20,LWHSZBCFZ22}. 
This phenomenon of accessing unintended information is referred to as attribute inference (\attrinf). 
State-of-the-art attacks often utilize embeddings from a specific sample ($\xtarget$) extracted from the model in question to ascertain the attributes of that sample. 
Thus, we assume the adversary to have white-box access to the target model.
Followed by previous work~\cite{LWHSZBCFZ22}, attribute inference is formally described as follows:
\begin{equation}
\begin{split}
\attrinf:\xtarget,\WM,\SD\rightarrow {\textit{target attributes}}
\end{split}
\end{equation}
Here, $\dset{\aux}{}$ represents an auxiliary dataset that contains a secondary attribute. 
It is assumed that the adversary has the ability to build the target attributes in the auxiliary dataset and employs the embeddings of these attributes from the auxiliary dataset to train a classifier, aiming to predict the attributes of the actual dataset.

%-------------------------------------------------------------------------------
\subsection{Membership Inference}
\label{section:meminf}
%-------------------------------------------------------------------------------

Membership inference attacks (\meminf)~\cite{SSSS17} involve adversaries seeking to ascertain if a specific data point was used in the training of a machine learning model. 
Specifically, given a data sample $\xtarget$, a target model $\model{}{}$, and an auxiliary dataset $\dset{\aux}{}$, the process of membership inference is formulated as:
\begin{equation}
\begin{split}
\meminf:\xtarget,\model{}{},\dset{\aux}{}\rightarrow {\textit{member}, \textit{non-member}}
\end{split}
\end{equation}
wherein $\model{}{}\in{\BM,\WM}$ and $\dset{\aux}{}\in{\PD,\SD}$.

Extensive papers have been conducted on membership inference~\cite{SSSS17,NSH18,SZHBFB19,JSBZG19,SDSOJ19,LZ21,CYZF20,LF20,CZWBHZ21,LWHSZBCFZ22}, emphasizing its potential to compromise privacy. 
For instance, if a model for predicting medication dosages uses data from patients with a specific ailment, the model’s training inclusion reveals sensitive health information. 
Predominantly, such inference attacks indicate that a model may reveal extra information, facilitating further exploits~\cite{C20}.

Below is an explanation of how to implement membership inference (\meminf) under varying threat scenarios.

\mypara{Black-Box/Shadow \atktu{\meminf}{\BM}{\SD}~\cite{SZHBFB19}}
The most prevalent and challenging scenario involves the adversary having only black-box access (\BM) to the model along with a shadow auxiliary dataset (\SD). 
The adversary divides the shadow dataset and trains a model similar to the target model (mostly with the same architecture) on the shadow training dataset.
For each sample, the output from this shadow model indicates membership, which the adversary uses to label the data.
After finishing the shadow training process, the adversary uses the shadow testing dataset to query the shadow model. 
The adversary labels these samples as non-member data.
These labeled data then help train a meta-classifier that determines membership in the target model by analyzing the output from the target model.

\mypara{Black-Box/Partial \atktu{\meminf}{\BM}{\PD}~\cite{SZHBFB19}}
When the adversary has black-box access and only partial data from the training dataset, they can strategically leverage these data samples to query the target model directly, eliminating the necessity of training a shadow model. 
The outputs from the target model with partial training data can be effectively categorized as the ground truth of members. 
More concretely, the adversary obtains non-member data by querying the target model using samples from a testing dataset. 
Upon acquiring these labeled datasets, the adversary can proceed to develop a meta-classifier.

\mypara{White-Box/Shadow \atktu{\meminf}{\WM}{\SD}\cite{NSH19}}
Nasr et al.~\cite{NSH19} introduce an attack in the white-box setting. 
Although their initial configuration utilized partial training datasets, we follow previous work~\cite{LWHSZBCFZ22} by extending this setting to incorporate shadow models.
Similar to the black-box model scenario, this setting necessitates the training of a shadow model to obtain ground truth for member and non-member samples. 
However, a crucial difference from the black-box setting lies in the assumption that the adversary has full access to the target models. 
This enhanced access enables the adversary to strengthen membership inference through the utilization of supplementary information. 
In this paper, we follow the methodology established by Nasr et al.~\cite{NSH19}, conducting membership inference attacks by leveraging multiple features: sample gradients concerning the model parameters, embeddings from different intermediate layers, classification loss, and prediction posteriors (and labels).

\mypara{White-Box/Partial \atktu{\meminf}{\WM}{\PD}~\cite{NSH19}}
The method in this scenario mirrors \atktu{\meminf}{\BM}{\PD}. 
The only difference is that the adversary can utilize the features of \atktu{\meminf}{\WM}{\SD}.

\mypara{LiRA \atktu{\meminf}{\LIRA}{\SD}~\cite{CCNSTT22}}
Unlike the previous four settings that rely on confidence scores or prediction probabilities, \LIRA leverages multiple shadow models and constructs a likelihood ratio based on the difference in model outputs when data points are included or excluded from training.
Therefore, we need to train different shadow models with different \SD.

%-------------------------------------------------------------------------------
\subsection{Property Inference}
\label{section:propinf}
%-------------------------------------------------------------------------------

Property inference attacks (\propinf)~\cite{MSCS19,GWYGB18,MGC22,ZCSZ22} aim to infer general information about the training dataset, such as the proportion of data with a specific property unrelated to the main classification task.
For example, the gender ratio in the training dataset can be inferred when a model for classifying race is given.
Previous works require access to the
training process of the model (e.g., via gradients~\cite{MSCS19}) or to model parameters~\cite{GWYGB18}.
These methods are easy to implement for a few layers of neural networks.
However, once the model becomes complex, the vast computational and memory resources are difficult to achieve. 
In addition, we build the query auxiliary datasets \dset{\aux}{Q} with different proportions of property.
Therefore, in this paper, given a target model $\model{}{}$, the adversary first trains the shadow models by shadow auxiliary datasets \SD with different proportions of the target property.
Next, they query these shadow models to get the outputs of each proportion and concatenate these results together to train a meta-classifier for the property inference.
We only need black-box access for this attack.
Thus, the property inference can be defined as:
\begin{equation}
\small
\begin{split}
    \propinf:\BM,\dset{\aux}{\mathsf{T}},\SD\rightarrow \{\textit{target property}\}
\end{split}
\end{equation}

The global properties of a dataset are confidential when they relate to the proprietary information or intellectual property that the data contains, which its owner is not willing to share.
This exposure can lead to severe privacy violations, especially if the data is protected by regulations like GDPR~\cite{GDPR}.

%-------------------------------------------------------------------------------
\section{Threat Modeling}
\label{section:threat}
%-------------------------------------------------------------------------------

This work focuses on image classification ML models, where the model takes a data sample as input and outputs a probability vector, known as posteriors. 
Each component of the posteriors represents the likelihood that the sample belongs to a specific class.

We categorize the threat models along two dimensions: 1) \textit{access to the target model} and 2) \textit{availability of an auxiliary dataset}.

\mypara{Access to the Target Model}
We consider two access settings: \textit{white-box} and \textit{black-box}. In the white-box setting (\WM), the adversary has full knowledge of the target model, including its parameters and architecture. In contrast, the black-box setting (\BM) limits the adversary to interact with the model like an API, where they can only query it and receive outputs. However, much of the black-box literature~\cite{SSSS17,GWYGB18,XWLBGL21} also assumes the adversary knows the model's architecture, which they use to build shadow models (see \autoref{section:preliminary}).

\mypara{Auxiliary Dataset}
The adversary needs an auxiliary dataset to train their attack model.
% Specifically, we consider three level
For this knowledge, we consider three scenarios: 1) \textit{partial training dataset} (\PD), 2) \textit{shadow auxiliary dataset} (\SD), and 3) \textit{query auxiliary dataset} (\QD). 
In the first scenario, the adversary acquires part of the real training data of the target model (datasets where it is public knowledge).
For the \SD setting, the adversary gets a ``shadow'' dataset from the same distribution as the training data of the target model, which is used to train a shadow model (see Section V-C in~\cite{SSSS17} for a discussion on how to generate such data).
In the last scenario, the adversary establishes a dataset with different property proportions to query the shadow model, thereby training the attack model for \propinf. 
This dataset is never used to train either the target model or the shadow model, and it needs to have the same distribution as the target training dataset.
Unlike the first two settings, \QD is constructed based on the second property proportions that may exist during model training (see \autoref{section:propinf}).

%-------------------------------------------------------------------------------
\section{Attack Composition}
\label{section:combomemethod}
%-------------------------------------------------------------------------------

In this section, we introduce our hierarchical compositions of different attack types. First, we propose a taxonomy that offers a structured framework for studying these compositions. Next, we outline the methodologies for specific attack compositions, designating one as the \textit{primary attack} and enhancing it with a \textit{support attack}.

%-------------------------------------------------------------------------------
\subsection{Attack Composition Taxonomy}
\label{section:combo}
%-------------------------------------------------------------------------------

To address \textbf{RQ1}, which examines the approaches for designing and implementing attack compositions, we propose a taxonomy based on the attack pipeline. 
This taxonomy serves several purposes: (1) Most attack pipelines consist of multiple phases, allowing integration and composition of different attacks at various phases.
(2) It is both domain- and model-agnostic, making it easily adaptable to other areas, such as graph data, NLP, and transformer-based models.
(3) It offers future researchers a clear framework for studying attack compositions, providing potential benefits to the community. 

\mypara{Preparatory Level}
In the preparation stage, the adversary gathers information, sets up the environment, and develops the necessary tools. 
This includes collecting data about the target machine learning system, such as input-output pairs, model parameters, and any accessible metadata, to understand its architecture. 
The adversary develops or selects appropriate attack algorithms, like FGSM~\cite{GSS15} in adversarial example attack, and sets up frameworks and libraries, like PyTorch~\cite{PyTorch} or CleverHans~\cite{PFCGFKXSBRMBHZJLSGUGDBHRLM18}. Additionally, the adversary prepares the computational infrastructure, including high-performance GPUs or cloud services, and may train a shadow/surrogate model to simulate the target system. 

\mypara{Execution Level}
During the execution phase, the actual attack is executed against the target machine learning system. 
For example, the adversary may deploy the attack by generating adversarial examples through perturbing input data to mislead target models or replicating the target model via model extraction. 
Throughout this phase, the adversary collects outputs and logs detailed data from the target system for subsequent analysis.

\mypara{Evaluation Level}
In the evaluation phase, the adversary analyzes the outcomes, assesses the attack performance, and identifies areas for improvement. 
This involves defining and measuring success metrics, such as misclassification rates or confidence reductions, and assessing the broader impact on system performance and security. 
Post-attack analysis includes examining the types of errors induced by the attack and studying changes in model behavior to understand vulnerabilities.
Insights gained during this phase guide the refinement and iteration of the attack strategy, enhancing its effectiveness in subsequent attempts.

%-------------------------------------------------------------------------------
\subsection{Preparation Level}
\label{section:preplevel}
%-------------------------------------------------------------------------------

We first introduce attack compositions at the preparatory stage. 
Here, the support attack assists the primary attack during preparation before the primary attack is executed.

%-------------------------------------------------------------------------------
\subsubsection{\atoa{\propinf}{\attrinf}}
%-------------------------------------------------------------------------------

The first attack composition is enhancing \attrinf (primary attack) by using \propinf (support attack) during its preparatory stage.
Specifically, adversaries in \attrinf often overlook a key issue: creating a more effective auxiliary dataset for training attack models. 
The target attribute bias of the target model's training dataset can complicate the auxiliary dataset, making it crucial to address this bias during preparation. 
Therefore, we amplify \attrinf by employing \propinf to assist in dataset construction during the preparatory phase.

In general, our intuition is that {\color{NavyBlue} \propinf can better assist in determining the proportion of the target attribute in the training dataset.}
For \attrinf, we believe that adversaries will not really care about the proportion of the target attribute in the auxiliary dataset.
They can never fully eliminate the influence of the bias in the target model without knowing the property information. 
Therefore, we first determine the distribution of the target attribute in the training dataset using \propinf and further sample the auxiliary dataset, significantly enhancing the effectiveness of \attrinf.
In general, the \atoa{\propinf}{\attrinf} can be defined as:
\begin{equation}
\begin{aligned}
    \atoa{\propinf}{\attrinf}&:\xtarget,\WM,\dset{\aux}{},\propinf\\
    &\quad\quad\quad\quad\quad\quad \rightarrow\{\textit{target attributes}\}
\end{aligned}
\end{equation}
More concretely, we have two different scenarios for utilizing \propinf, i.e., empirical and theoretical settings.
1) For the empirical setting, we use the real posterior of the \propinf attack model as the confidence for sampling the \attrinf training dataset.
For the proportion of the property $p$, given the confidence $c$, the ratio of sampling is $c \times (1-p)$.
2) On the other hand, for the theoretical setting, we directly use the predicted label from \propinf into the sampling function.
In general, when enough shadow models are trained, such as 1,000 for each label, the empirical setting becomes the theoretical setting.

%-------------------------------------------------------------------------------
\subsection{Execution Level}
\label{section:execlevel}
%-------------------------------------------------------------------------------

At the execution level, the support attack interacts simultaneously with the primary attack during its execution. 
This concurrent interaction can amplify the impact of the primary attack by leveraging the synergistic effects of support attacks.

%-------------------------------------------------------------------------------
\subsubsection{\atoa{\adv}{\meminf}}
%-------------------------------------------------------------------------------

Previous work~\cite{LZ21} has demonstrated {\color{NavyBlue} a distribution shift between the members and non-members when calculating the distance between the adversarial examples and the original images.}
Following this intuition, we trade this distance as additional information to assist \meminf.
For the \atktu{\meminf}{\BM}{\dset{\aux}{}}, we choose a black-box adversarial attacks, $\mathsf{Square}$~\cite{ACFH20}.
$\mathsf{Square}$ is a score-based black-box adversarial attack that does not rely on a local gradient.
Instead, it utilizes a randomized search scheme that selects localized square-shaped updates at random positions so that at each iteration, the perturbation is situated approximately at the boundary of the dataset.
For the \atktu{\meminf}{\WM}{\dset{\aux}{}}, we choose a white-box adversarial attack, $\mathsf{PGD}$~\cite{MMSTV18}.
It is an iterative method that makes small modifications to the input data at each step by computing the gradient of the loss function with respect to the input data. 
This gradient demonstrates how to change the input slightly to increase the loss.
When the noise $\delta$ added by the $\mathsf{Square}$ or $\mathsf{PGD}$ is able to change the prediction of the original label, we stop adding noise and use the data $\xattack{adv} = \xtarget + \delta$ as adversarial examples.
In the experiments, we find that calculating the norm of the distance provides better assistance than directly using the distance.
For \LIRA, we estimate the joint distribution of members and non-members using their logits and auxiliary scores.
Based on the shadow model outputs, we calculate the covariance matrix and mean separately for members and non-members.
For each target sample, we compute the log-likelihood under the member and non-member distributions using the multivariate normal PDF.
The LiRA score is computed as the difference between the two log-likelihoods: $\text{score} = -\log P(x \mid \text{in}) + \log P(x \mid \text{out})$.

Therefore, we first calculate the $L_{2}$ distance between member (non-member) samples and their adversarial samples in the auxiliary dataset \dset{\aux}{}.
Next, in addition to the normal inputs required for \meminf, such as outputs from the target or shadow model and predicted labels, we also use the $L_{2}$ distances as other inputs to train the attack model.
As a result, \atoa{\adv}{\meminf} can be defined as:
\begin{equation}
\begin{aligned}
    \atoa{\adv}{\meminf}&:\xtarget,\model{}{},\dset{\aux}{},L_{2}^{\dset{\aux}{}} \\
    &\quad\quad\quad\quad \rightarrow\{\textit{member}, \textit{non-member}\}
\end{aligned}
\end{equation}

%-------------------------------------------------------------------------------
\subsubsection{\atoa{\adv}{\propinf}}
%-------------------------------------------------------------------------------

Currently, \propinf heavily depends on training a large number of shadow models. 
The more shadow models, the better the effectiveness of \propinf. 
However, training such a large number of shadow models is computationally expensive. 
Therefore, we hope to find additional information to reduce the number of shadow models and increase the accuracy of \propinf.
Thus, similar to \atoa{\adv}{\meminf}, our intuition is, {\color{NavyBlue} for the auxiliary datasets $\dset{\aux}{\mathsf{T}}$ with different proportions of the target property, the distribution of the $L_{2}$ distance between these samples and their adversarial samples should also be different.}
For example, the distributions of $L_{2}$ distances calculated on the auxiliary dataset by models trained on a male-to-female ratio of 5:5 versus 2:8 are different.
Following this intuition, we concatenate these $L_{2}$ distances with the original inputs of \propinf together to train a meta-classifier.
\atoa{\adv}{\propinf} can be defined as:
\begin{equation}
\begin{aligned}
    \atoa{\adv}{\propinf}&:\model{}{},\QD,\SD,L_{2}^{\QD}\\
    &\quad\quad\quad\quad\quad\quad \rightarrow \{\textit{target property}\}
\end{aligned}
\end{equation}

%-------------------------------------------------------------------------------
\subsection{Evaluation Level}
\label{section:evallevel}
%-------------------------------------------------------------------------------

In the evaluation stage, the support attack assists the primary attack after its initial execution. 
This post-attack support can refine the primary attack's outcomes, correct discrepancies, or further exploit vulnerabilities. 
In other words, the support attack serves to \textit{calibrate} the results of the primary attack.

%-------------------------------------------------------------------------------
\subsubsection{\atoa{\propinf}{\meminf}}
%-------------------------------------------------------------------------------

Previous work~\cite{ZCSZ22} finds that \propinf on GAN models can improve the effectiveness of \meminf. 
\meminf is enhanced by calibrating the output of the attack model with the proportion of the target property $\lambda_{p}\frac1N\sum_i^N(\mathcal{P}_i-0.5)$.
Among those, $\lambda_{p}$ controls the magnitude of the enhancement.
$\mathcal{P}_{i}-0.5$ is the proportion of the label to which the target sample belongs.
However, for ML models, this calibration is equivalent to directly finding another threshold to classify \meminf.
In this scenario, our intuition is {\color{NavyBlue} a sample has a larger possibility of being a member when it shares the same property with most samples in the target property.}
Unlike previous work~\cite{ZCSZ22}, we further train an encoder $\mathcal{E}$ to select different $\lambda$s for the calibration during the attack model training phase, thereby boosting \meminf more effectively.
Note that the input of the encoder is the output of the target model \model{}{}.
Formally, the new calibration of \meminf is defined as:
\begin{equation}
\begin{aligned}
    \atoa{\propinf}{\meminf}&:\xtarget,\model{}{},\dset{\aux}{},\lambda\\
    &\quad\quad \rightarrow \{\textit{member}, \textit{non-member}\}
\end{aligned}
\end{equation}
For normal \meminf, $\lambda$ is a set of $\mathcal{E}(\model{}{}(\dset{\aux}{}))$ and the calibration function is $\lambda\frac1N\sum_i^N(\mathcal{P}_i-0.5)$.
Since \propinf in our scenario is a black-box attack, we can relax this information on both black/white-box \meminf.
Specifically, different from \atoa{\propinf}{\attrinf}, since the confidence of \propinf in this scenario is a constant number, there is no difference between empirical and theoretical settings.
For \LIRA, when we know the prior distribution of a certain property (e.g., class label, gender, category) in the target model training dataset, we can incorporate this knowledge to refine the \LIRA membership inference scores. 
Specifically, we adjust the \LIRA score of each sample based on the prior probability of its associated property.
Let $s_i$ be the original \LIRA score for sample $i$, and let $p_i$ be the property value associated with that sample. 
If the prior distribution over the property is known and denoted as $P(p)$, we can compute a prior-adjusted score: $\lambda = s_i \cdot P(p_i)$.

%-------------------------------------------------------------------------------
\section{The \systemname Toolkit}
\label{section:module}
%-------------------------------------------------------------------------------

In this section, we present \systemname, a modular toolkit designed to evaluate the above attack compositions.
Researchers have developed several software tools to measure the potential security/privacy risks of ML models, such as DEEPSEC~\cite{LJZWWLW19} and CleverHans~\cite{PFCGFKXSBRMBHZJLSGUGDBHRLM18} for evaluating adversarial example attacks, TROJANZOO~\cite{PZGXJCW20} for backdoor attacks, and ML-Doctor~\cite{LWHSZBCFZ22} for jointly analyzing the relationships among different attacks.
Inspired by this work, we design a systematic framework to modularize our experiments better, namely \systemname.
To our knowledge, \systemname is the first framework that jointly considers the composition of different inference-time attacks.

\begin{figure}
\centering
\includegraphics[width=1.0\columnwidth]{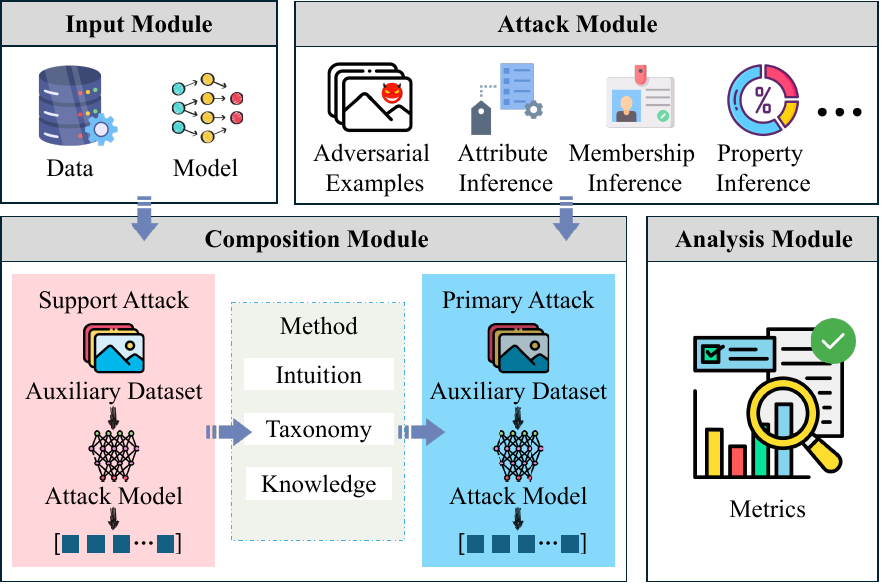}
\caption{Overview of the workflow of \systemname. }
\label{figure:workflow}
\end{figure}

\mypara{Modules} \autoref{figure:workflow} illustrates the four modules of \systemname:

\begin{enumerate}
\item \textbf{Input.} 
This module prepares the dataset and model for the other modules.
More precisely, it performs dataset partition/preprocessing, constructs model architectures, and trains the model.
\item \textbf{Attack.} 
This module includes four inference-time attacks, each employing the most representative strategy. 
These attacks can be seamlessly replaced or updated with newer versions.
\item \textbf{Composition.} 
This module implements attack compositions where one support attack assists a primary attack.
Currently, we have introduced four specific attack composition methods. 
Notably, users can add new composition methods as needed.
\item \textbf{Analysis.}
This module evaluates and compares the performance of individual attacks and attack compositions. 
We include various evaluation metrics to provide a comprehensive analysis.
\end{enumerate}

Overall, the modular design of \systemname allows researchers and practitioners to reuse it as a standard benchmark tool, experimenting with new and additional datasets, model architectures, and attacks.

%-------------------------------------------------------------------------------
\section{Experimental Settings}
\label{section:expset}
%-------------------------------------------------------------------------------

We first select three benchmark datasets (see \autoref{section:dataset}) and three state-of-the-art ML models (see \autoref{section:target_model}) to train thousands of target and shadow models.
For each dataset, we partition it into four parts (see \autoref{section:dataset}), including the target training dataset, target testing dataset, shadow training dataset, and shadow testing dataset, to comply with the different scenarios discussed in \autoref{section:combo}.

%-------------------------------------------------------------------------------
\subsection{Datasets}
\label{section:dataset}
%-------------------------------------------------------------------------------

In this work, we consider three benchmark datasets.

\begin{itemize}
\item \textbf{CelebA~\cite{LLWT15}} contains 202,599 face images, each labeled with 40 binary attributes. 
We select three attributes—\emph{HighCheekbones}, \emph{WearingNecktie}, and \emph{ArchedEyebrows}—to define the target models' classes. The first two attributes form a 4-class classification for the first property, while the third attribute represents the second property.
\item \textbf{CIFAR10~\cite{CIFAR}} is a widely used dataset containing 60,000 32x32 color images across ten classes, with 6,000 images per class. We group the second property into two categories: animal and non-animal.
\item \textbf{Places~\cite{ZLKOT18}} contains 1.8 million training images from 365 scene categories. The validation set has 50 images per category, and the test set has 900. For our study, we select 20 scenes, with 3,000 images each, and group them into two categories—indoor and outdoor—for the second property.
\end{itemize}

We partition each dataset into four components to facilitate comprehensive evaluation. 
The first part comprises the target training dataset. 
For \propinf, we employ different random seeds to select samples based on the second property, ensuring alignment with the desired proportional distribution. 
For other configurations, we maintain the balanced proportional distribution in the original dataset. 
The second part constitutes the target test dataset, which is methodically balanced across various properties to ensure unbiased evaluation. 
The third part encompasses the shadow training dataset, which is constructed following the same method utilized for the target training dataset. 
Finally, the fourth part consists of the shadow test dataset, which mirrors the selection criteria applied to the target test dataset, maintaining consistency in our experimental framework.
Note that this dataset splitting is the basic setup in this field~\cite{SSSS17,NSH18,SZHBFB19,LWHSZBCFZ22,HLXCZ22,LLHYBZ222,LZBZ222,FWGLLJ23}.

%-------------------------------------------------------------------------------
\subsection{Target Models}
\label{section:target_model}
%-------------------------------------------------------------------------------

We select three widely-used ML models, i.e., DenseNet121~\cite{HLMW17}, ResNet18~\cite{HZRS16}, and VGG19~\cite{SZ15}.
We set the mini-batch size to 256 and use cross-entropy as the loss function.
We use Adam~\cite{KB15} as the optimizer with a learning rate of 1e-2.
Each target model is trained for at most 100 epochs.
Once the overfitting is larger than 0.250, model training is finished.
Note that for shadow models used in the \meminf and \propinf, we train thousands following the same process as the target models with the support of SecurityNet~\cite{ZLYHBFZ24}.

%-------------------------------------------------------------------------------
\subsection{Attack Models}
\label{section:attack_model}
%-------------------------------------------------------------------------------

\mypara{Attribute Inference} 
At the preparatory level, the assistant from \propinf will not influence the types of inputs.
Therefore, our attack model is a 2-layer MLP where its input is the embeddings from the second-to-last layer of the target model. 
We use cross-entropy as the loss function and Adam as the optimizer with a learning rate of 1e-2.
The attack model is trained for 100 epochs.
We use {\em accuracy} and {\em F1 score} for the evaluation metrics.

\begin{table*}[!t]
\centering
\caption{Performance of target models, namely, training/testing accuracy for each setting.
We also provide the results of different proportions of the second property.}
\label{table:model_acc}
\setlength{\tabcolsep}{4 pt}
\customTableFont
\begin{tabular}{@{}l| c c c c c c@{}}
\toprule
& \multicolumn{2}{c}{\bf CelebA} & \multicolumn{2}{c}{\bf CIFAR10} & \multicolumn{2}{c}{\bf Places} \\
Property Proportion & 2:8 & 5:5 & 2:8 & 5:5 & 2:8 & 5:5 \\
\midrule
{\bf DenseNet121} & 0.988/0.835 & 0.987/0.840 & 0.866/0.653 & 0.882/0.687 & 0.844/0.634 & 0.883/0.668 \\
{\bf ResNet18} & 0.994/0.829 & 0.993/0.834 & 0.812/0.600 & 0.896/0.677 & 0.821/0.584 & 0.709/0.589 \\
{\bf VGG19} & 0.935/0.833 & 0.937/0.845 & 0.764/0.565 & 0.843/0.645 & 0.842/0.668 & 0.878/0.677 \\
\bottomrule
\end{tabular}
\end{table*}

\begin{table*}[!t]
\centering
\caption{Performance of \atoa{\propinf}{\attrinf}.
Here, the empirical setting is based on the confidence (posterior) of \propinf, while the theoretical setting is the label of the prediction of \propinf.}
\label{table:propinf2attrinf}
\setlength{\tabcolsep}{4 pt}
\customTableFont
\begin{tabular}{@{}l l | c c c c c c@{}}
\toprule
& & \multicolumn{2}{c}{\bf CelebA} & \multicolumn{2}{c}{\bf CIFAR10} & \multicolumn{2}{c}{\bf Places} \\
{\bf Model} & {\bf Mode} & {\bf F1 Score} & {\bf Accuracy} & {\bf F1 Score} & {\bf Accuracy} & {\bf F1 Score} & {\bf Accuracy} \\
\midrule
\multirow{3}{*}{\bf DenseNet121} & Origin & 0.771 & 0.712 & 0.916 & 0.911 & 0.667 & 0.500 \\
& Empirical & 0.789 & 0.780 & 0.930 & 0.929 & 0.923 & 0.921 \\
& Theoretical & 0.782 & 0.783 & 0.930 & 0.930 & 0.916 & 0.914 \\
\midrule
\multirow{3}{*}{\bf ResNet18} & Origin & 0.779 & 0.736 & 0.667 & 0.500 & 0.667 & 0.500 \\
& Empirical & 0.790 & 0.772 & 0.895 & 0.894 & 0.901 & 0.895 \\
& Theoretical & 0.789 & 0.774 & 0.880 & 0.872 & 0.911 & 0.909 \\
\midrule
\multirow{3}{*}{\bf VGG19} & Origin & 0.742 & 0.664 & 0.911 & 0.905 & 0.915 & 0.910 \\
& Empirical & 0.757 & 0.747 & 0.918 & 0.921 & 0.937 & 0.937 \\
& Theoretical & 0.759 & 0.748 & 0.917 & 0.917 & 0.937 & 0.937 \\
\bottomrule
\end{tabular}
\end{table*}

\mypara{Membership Inference}
Recall that there are four different scenarios for \meminf; we establish two types of attack models: one for the black-box and the other for the white-box setting.
For black-box settings, our original attack model has two inputs: the target sample's ranked posteriors and a binary indicator on whether the target sample is predicted correctly. 
Each input is first fed into a different 2-layer MLP. 
Then, the two obtained embeddings are concatenated and fed into a 4-layer MLP.
For the white-box, we have four inputs for this attack model, including the target sample's ranked posteriors, classification loss, gradients of the parameters of the target model's last layer, and one-hot encoding of its true label.
Each input is fed into a different neural network, and the resulting embeddings are concatenated as input to a 4-layer MLP.
We use ReLU as the activation function for the attack models.
For the attack scenario assisted by \adv, the inputs of both the black-box and white-box attack models expand the $L_{2}$ distance between each image and its adversarial example in the auxiliary dataset.
The original attack model remains the same for the attack scenario assisted by \propinf, but the encoder for choosing $\lambda$ is a 4-layer MLP.
The attack model is trained for 50 epochs by using the Adam optimizer with a learning rate of 1e-5.
We adopt {\em accuracy}, {\em F1 score}, {\em AUC score}, and {\em TPR @0.1\% FPR} as the evaluation metrics.

\mypara{Property Inference}
Recall that the algorithm level needs to add additional information during the attack phase.
For \propinf, the attack model is a meta-classifier; its inputs are organized from the unified overall outputs of each target (shadow) model by feeding the test auxiliary dataset with different proportions of another property.
For the assisted \propinf, the inputs also expand a one-dimensional vector composition of the $L_{2}$ distance between each image and its adversarial example in the test auxiliary dataset.
We adopt {\em accuracy} as the evaluation metric on 100 models.

%-------------------------------------------------------------------------------
\section{Experimental Evaluation} 
\label{section:evaluation}
%------------------------------------------------------------------------------

%-------------------------------------------------------------------------------
\subsection{Target Model Utility} 
\label{section:utility}
%------------------------------------------------------------------------------

First, we present target model utilities in \autoref{table:model_acc}. 
Based on previous work~\cite{LWHSZBCFZ22}, we define an overfitting level as the difference between its accuracy on the training and test datasets; the greater this difference, the more overfitting the model is. 
As shown, the overfitting levels in our target models are less than 0.250.
On the other hand, we ensure a real-world scenario as much as possible to validate the effectiveness of our attack composition.
Note that target models trained on datasets with a 2:8 proportion for the second property are used for \propinf, while a 5:5 proportion is used for other attacks.

\begin{figure*}[!t]
\centering
\includegraphics[width=2.0\columnwidth]{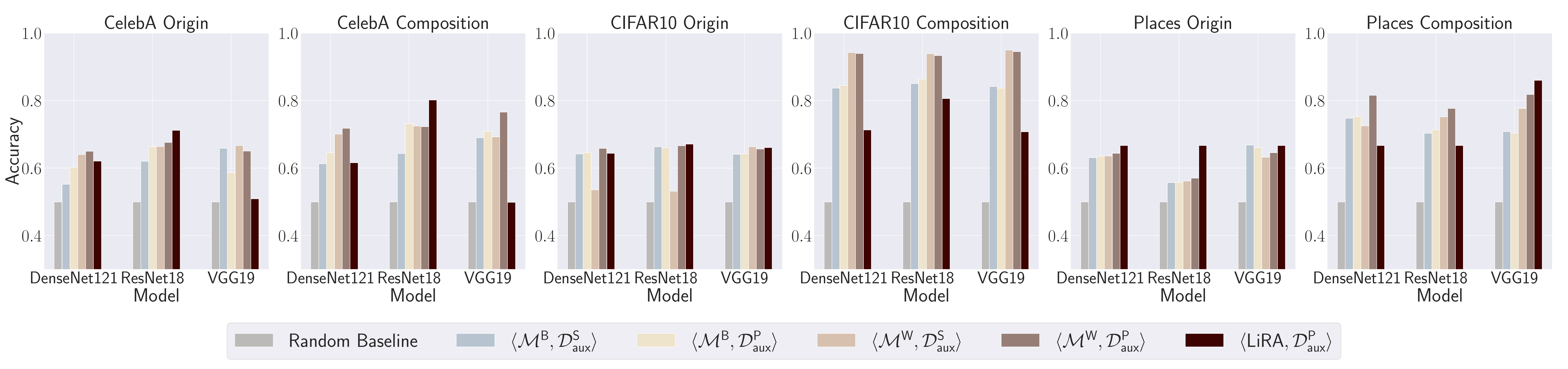}
\caption{Accuracy of \atoa{\adv}{\meminf} under different threat models, datasets, and target model architectures.}
\label{figure:adv2meminf_acc}
\end{figure*}

%-------------------------------------------------------------------------------
\subsection{Preparation Level} 
\label{section:preeva}
%------------------------------------------------------------------------------

At this level, since we only need to change the data preprocessing phase, the subsequent training of the attack model will remain consistent with the original attack. 
In this case, our focus will be on preprocessing the dataset. 
As mentioned before, we demonstrate this attack level through \atoa{\propinf}{\attrinf}.

%-------------------------------------------------------------------------------
\subsubsection{\atoa{\propinf}{\attrinf}}
%-------------------------------------------------------------------------------

We present the performance of \atoa{\propinf}{\attrinf} by comparing it with the original \attrinf first.
\autoref{table:propinf2attrinf} demonstrates the results of \atoa{\propinf}{\attrinf}.
We can find that the original \attrinf achieves a random guess for three scenarios.
This indicates that simply collecting datasets will easily cause severe bias in property proportions, making original \attrinf challenging to achieve. 
Besides, the results are obviously better than the original attacks in both empirical and theoretical settings.
For example, when using CIFAR10 to launch \attrinf on the DenseNet121 model, the original F1 score is 0.916, and accuracy is 0.911, while \atoa{\propinf}{\attrinf} can achieve 0.930 and 0.929 for the empirical setting as well as 0.930 and 0.930 for the theoretical setting.
This also means that with the assistance of \propinf, \attrinf can indeed achieve better results, which verifies our intuition: \propinf can better assist in determining the proportion of the target attribute in the original training dataset.

In addition, by training the \propinf attack model with 1,000 shadow models, the confidence of our target models exceeds 0.950.
Therefore, there is little essential difference between our empirical and theoretical settings.
In a nutshell, preprocessing in the preparatory phase is very intuitive, which requires us to choose a good assistant to complete.

%-------------------------------------------------------------------------------
\subsection{Execution Level} 
\label{section:execeval}
%------------------------------------------------------------------------------

At this level, we leverage \adv to assist two different types of attacks, \meminf and \propinf, during their execution stages.

%-------------------------------------------------------------------------------
\subsubsection{\atoa{\adv}{\meminf}}
%-------------------------------------------------------------------------------

First, we evaluate the results of \meminf. 
We report the accuracy in \autoref{figure:adv2meminf_acc} of \atoa{\adv}{\meminf}, 
while \autoref{figure:adv2meminf_f1}, \autoref{figure:adv2meminf_auc}, and \autoref{table:tpradv2meminf}, respectively, F1, AUC score, and TPR @0.1\% FPR.
For some experiments, the original attacks do not achieve much higher attack performance than the random baseline, which means that overfitting does not have a significant impact on the attack~\cite{SSSS17}; see \autoref{section:meminf}.
For instance, the original attack accuracy, F1 score, and AUC score of \atktu{\meminf}{\WM}{\PD} on ResNet18 trained on Places are 0.544, 0.572, and 0.570, respectively.
TPR @0.1\% FPR score is 0.001, which is very low in this scenario.
Compared to the previous works~\cite{CYZF20,LF20,CZWBHZ21}, white-box attacks have not significantly surpassed black-box attacks.
This is expected because, in these works, the training accuracy of the target model can reach 1.000, meaning that for the training dataset, i.e., members, their loss is very close to zero. 
Nevertheless, this is not the case for non-members, allowing \meminf to achieve a high success rate.
In contrast, since the training set accuracy does not reach 1.000 in our work, the loss may act as a form of noise in white-box attacks.
We emphasize that our setting is more in line with real-world scenarios.

On the other hand, we find that \adv indeed significantly improves \meminf.
For example, the composition attack accuracy, F1 score, and AUC score of \atktu{\meminf}{\WM}{\PD} on ResNet18 trained on Places is 0.743, 0.653, 0.777, improved by nearly 0.200 compared to the original \meminf.
TPR @0.1\% FPR score is also up to 0.490, indicating that our composition attack model is effective at identifying true positives, even under very conservative conditions.
More specifically, for the CelebA dataset, since we created a 4-class problem by combining the two labels of the first attribute, the ADV might not perform as well as on the other two datasets. 
This is because when noise affects one of the labels, it can change the combined class of the image, but this noise may not impact all the labels, leading to a smaller distance between members and non-members compared to the previous datasets.
In general, the result first confirms our intuition; there is a distribution shift between the members and non-members when calculating the distance between the adversarial examples and the original data samples.
In addition, for \adv, we believe that this distance has magnified the gap between members and non-members, resulting in an enhanced \meminf with a higher success rate.
Therefore, the above results verify our intuition: there is a distribution shift between the members and non-members when calculating the distance between the adversarial examples and the original images.

\begin{table*}[!t]
\centering
\caption{Performance of \atoa{\adv}{\propinf}.}
\label{table:adv2propinf}
\setlength{\tabcolsep}{3 pt}
\customTableFont
\begin{tabular}{@{}l| c c c c c c@{}}
\toprule
& \multicolumn{2}{c}{\bf CelebA} & \multicolumn{2}{c}{\bf CIFAR10} & \multicolumn{2}{c}{\bf Places} \\
{\bf Model} & {\bf Origin} & {\bf Composition} & {\bf Origin} & {\bf Composition} & {\bf Origin} & {\bf Composition} \\
\midrule
{\bf DenseNet121} & 0.520 & 0.600 & 0.850 & 0.910 & 0.620 & 0.750 \\
{\bf ResNet18} & 0.510 & 0.600 & 0.890 & 0.960 & 0.750 & 0.830 \\
{\bf VGG19} & 0.540 & 0.630 & 0.860 & 0.930 & 0.730 & 0.750 \\
\bottomrule
\end{tabular}
\end{table*}

%-------------------------------------------------------------------------------
\subsubsection{\atoa{\adv}{\propinf}}
%-------------------------------------------------------------------------------

Next, we report our experimental results of \atoa{\adv}{\propinf} in \autoref{table:adv2propinf}.
We can clearly see that with the assistance of \adv, \propinf is significantly improved, which confirms our previous intuition.
For example, the original \propinf on ResNet18 trained by CIFAR10 is 0.890 when using 100 shadow models.
Nevertheless, after the assistance of \adv, the accuracy is increased to 0.960, equivalent to saving the time required to train at least 300 extra shadow models.
Overall, the results of \atoa{\adv}{\propinf} verify our intuition: for the auxiliary datasets with different proportions of the target property, the distribution of the $L_{2}$ distance between these samples and their adversarial samples should also be different.

%-------------------------------------------------------------------------------
\subsection{Evaluation Level} 
\label{section:evaleval}
%------------------------------------------------------------------------------

At this stage, the support attack calibrates the results of the primary attack. 
In this work, we introduce \propinf to calibrate \meminf.

\begin{figure*}[!t]
\centering
\includegraphics[width=2.0\columnwidth]{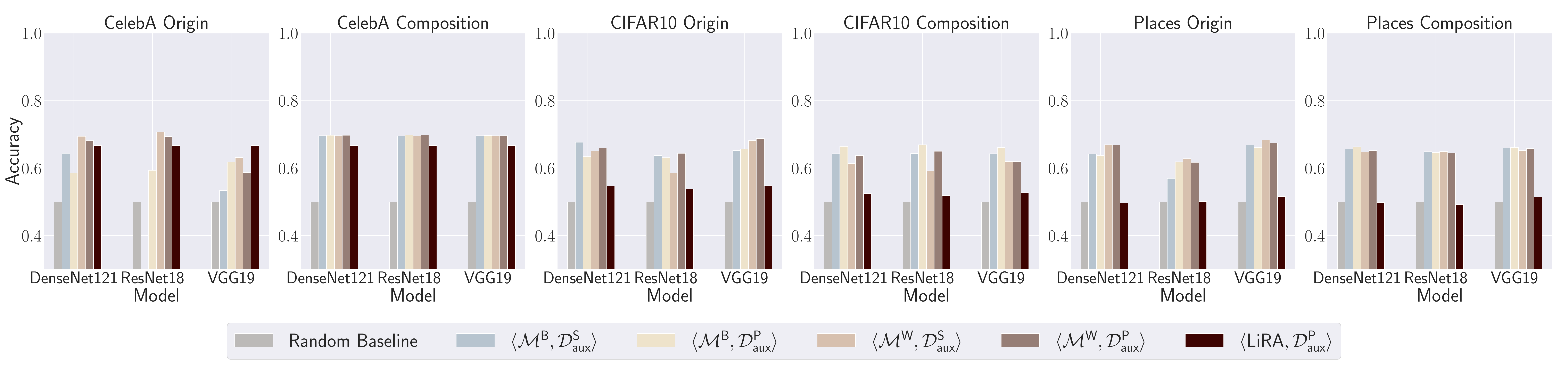}
\caption{Accuracy of \atoa{\propinf}{\meminf} under different threat models, datasets, and target model architectures.}
\label{figure:propinf2meminf_acc}
\end{figure*}

%-------------------------------------------------------------------------------
\subsubsection{\atoa{\propinf}{\meminf}}
%-------------------------------------------------------------------------------

We report the accuracy of \atoa{\propinf}{\meminf} in \autoref{figure:propinf2meminf_acc}.
We also report F1 score and AUC score (\autoref{figure:propinf2meminf_f1} and ~\autoref{figure:propinf2meminf_auc}) and, in \autoref{table:tprpropinf2meminf}, the TPR @0.1\% FPR results.
In many cases, the assistance of \propinf slightly improves \meminf's accuracy. 
While most TPR @0.1\% FPR values remain near zero, there are instances where the composition attack achieves a higher TPR. 
For example, the composition attack on ResNet18 trained on CIFAR10 shows an accuracy of 0.669, F1 score of 0.731, and AUC of 0.656, compared to the original 0.631, 0.695, and 0.617. 
The TPR @0.1\% FPR improves from 0.000 to 0.002. However, not all results show significant improvement.
We attribute this to the general nature of the information from \propinf, which lacks the detailed insights that \adv provides for training the entire model. 
Without rich data or clear distinctions between members and non-members, improvements in metrics like F1 score and AUC are limited, suggesting that the original \meminf may already be near its upper bound.
We attribute this to the general nature of the information from \propinf, which lacks the detailed insights that \adv provides for training the entire model. 
Improvements in metrics like F1 score and AUC are limited, suggesting that the original \meminf may already be near its upper bound.
We also observe that with \propinf's support, attack performance remains stable across different scenarios (black-box and white-box), indicating that \propinf helps \meminf approach its performance limit. 
These results confirm our intuition: a sample is more likely to be a member if it shares properties with most samples in the target group.

%-------------------------------------------------------------------------------
\subsection{Takeaways} 
\label{section:takeaway}
%-------------------------------------------------------------------------------

Overall, our evaluations demonstrate that combining different attack types significantly improves the effectiveness of primary attacks, leading to higher accuracy and success rates. 
These results confirm our earlier intuition about the benefits of attack compositions. 
Specifically, using \adv to assist \meminf and \propinf, as well as \propinf to assist \attrinf and \meminf, notably amplifies the ability to identify training data and infer sensitive information.
Our \systemname emerges as a valuable tool for systematically evaluating these compositions, highlighting the necessity for more robust defense mechanisms to counteract these amplified threats.

%-------------------------------------------------------------------------------
\section{Ablation Study} 
\label{section:ablation}
%------------------------------------------------------------------------------

%-------------------------------------------------------------------------------
\subsection{Differential Privacy}
\label{section:dp}
%-------------------------------------------------------------------------------

Differential privacy (DP)~\cite{DR14,LLSY16} is a mathematical framework for quantifying and protecting individual privacy in statistical analysis and machine learning. 
It formalizes privacy guarantees by ensuring that the inclusion or exclusion of any single data point in a dataset does not significantly affect the outcome of any analysis, thereby limiting the risk of re-identification.
DP has been widely adopted as a principled defense mechanism against membership inference attacks~\cite{NSH18,JE19,HZ21,LWHSZBCFZ22,NSTPC21}, offering formal privacy guarantees to mitigate the risk of exposing individual data records.
However, to our best knowledge, DP is not a defense mechanism against other attacks.
In this paper, we incorporate DP into the model training process by applying DP mechanisms to the optimizer, specifically using the Adam optimizer as mentioned in \autoref{section:target_model}. 
After training, we systematically evaluate the impact of differential privacy on the attack compositions by using \systemname. 

\mypara{Model Performance}
We first present the model performance with three different $\epsilon$ values on \autoref{table:model_acc_dp}.
Since adding DP into models requires a large computation cost, we conduct an in-depth research of the impact of DP on attack composition using the ResNet architecture as our main model.
From the table, only CelebA can achieve a standard performance without large utility degradation.
For the other two datasets, DP largely influences the performance.

\begin{table}[!t]
\centering
\caption{Performance of DP target models, namely, training/testing accuracy for each setting.
We also provide the results of different proportions of the second property.
Specifically, $\delta=1e-05$.}
\label{table:model_acc_dp}
\setlength{\tabcolsep}{4 pt}
\customTableFont
\begin{tabular}{@{}c c | c c c@{}}
\toprule
\multicolumn{2}{c}{\bf ResNet18} & {\bf CelebA} & {\bf CIFAR10} & {\bf Places} \\
\midrule
\multirow{2}{*}{\bf $\epsilon = 10$} & 2:8 & 0.831/0.730 & 0.248/0.286 & 0.268/0.192 \\
& 5:5 & 0.772/0.762 & 0.271/0.327 & 0.223/0.233 \\
\midrule
\multirow{2}{*}{\bf $\epsilon = 20$} & 2:8 & 0.841/0.753 & 0.275/0.340 & 0.304/0.207 \\
& 5:5 & 0.785/0.767 & 0.336/0.402 & 0.255/0.249 \\
\midrule
\multirow{2}{*}{\bf $\epsilon = 50$} & 2:8 & 0.851/0.754 & 0.346/0.415 & 0.377/0.251 \\
& 5:5 & 0.866/0.740 & 0.389/0.456 & 0.304/0.295 \\
\bottomrule
\end{tabular}
\end{table}

\begin{table*}[!t]
\centering
\caption{Performance of \atoa{\propinf}{\attrinf} with adding DP.
Here, the empirical setting is based on the confidence (posterior) of \propinf, while the theoretical setting is the label of the prediction of \propinf.}
\label{table:propinf2attrinf_dp}
\setlength{\tabcolsep}{4 pt}
\customTableFont
\begin{tabular}{@{}l l | c c c c c c@{}}
\toprule
& & \multicolumn{2}{c}{\bf CelebA} & \multicolumn{2}{c}{\bf CIFAR10} & \multicolumn{2}{c}{\bf Places} \\
& {\bf Mode} & {\bf F1 Score} & {\bf Accuracy} & {\bf F1 Score} & {\bf Accuracy} & {\bf F1 Score} & {\bf Accuracy} \\
\midrule
\multirow{3}{*}{\bf $\epsilon = 10$} & Origin & 0.667 & 0.500 & 0.667 & 0.500 & 0.667 & 0.500 \\
& Empirical & 0.776 & 0.769 & 0.742 & 0.743 & 0.805 & 0.829 \\
& Theoretical & 0.778 & 0.769 & 0.734 & 0.744 & 0.806 & 0.828 \\
\midrule
\multirow{3}{*}{\bf $\epsilon = 20$} & Origin & 0.667 & 0.500 & 0.667 & 0.500 & 0.667 & 0.500 \\
& Empirical & 0.777 & 0.772 & 0.790 & 0.786 & 0.854 & 0.880 \\
& Theoretical & 0.770 & 0.774 & 0.785 & 0.787 & 0.854 & 0.878 \\
\midrule
\multirow{3}{*}{\bf $\epsilon = 50$} & Origin & 0.667 & 0.500 & 0.667 & 0.500 & 0.667 & 0.500 \\
& Empirical & 0.787 & 0.778 & 0.759 & 0.785 & 0.813 & 0.825 \\
& Theoretical & 0.783 & 0.778 & 0.761 & 0.794 & 0.820 & 0.836 \\
\bottomrule
\end{tabular}
\end{table*}

%-------------------------------------------------------------------------------
\subsubsection{\atoa{\propinf}{\attrinf}}
%-------------------------------------------------------------------------------

\autoref{table:propinf2attrinf_dp} shows the results of \atoa{\propinf}{\attrinf} when DP is added.
First, DP does not impact \propinf, thus we can still achieve robust empirical results. 
Consequently, regarding the \atoa{\propinf}{\attrinf}, DP proves even less effective as a defense mechanism against composition.

%-------------------------------------------------------------------------------
\subsubsection{\atoa{\adv}{\meminf}}
%-------------------------------------------------------------------------------

\begin{figure*}[!t]
\centering
\includegraphics[width=2.0\columnwidth]{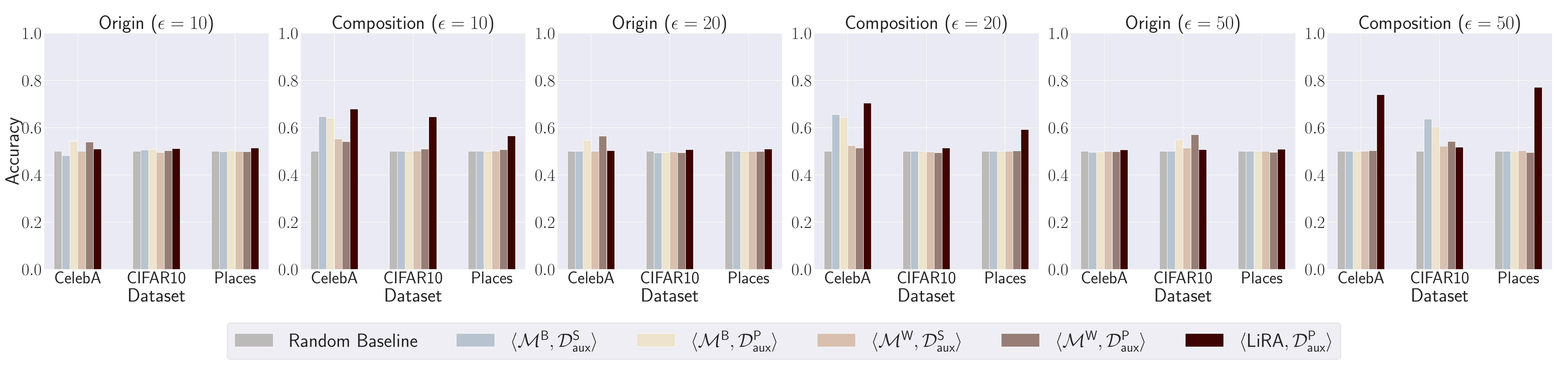}
\caption{Accuracy of \atoa{\adv}{\meminf} under different threat models, datasets, and target model architectures with adding DP.}
\label{figure:adv2meminf_acc_dp}
\end{figure*}

\autoref{figure:adv2meminf_acc_dp}, \autoref{figure:adv2meminf_f1_dp}, \autoref{figure:adv2meminf_auc_dp}, and \autoref{table:tpradv2meminf_dp} illustrate the effectiveness of \atoa{\adv}{\meminf} when DP is implemented during the training phase. 
The results indicate that although DP generally serves as an effective defense mechanism against \meminf attacks, it can be partially bypassed when adversarial examples are incorporated. 
For instance, in the case of \tuple{\BM, \SD} on the CIFAR10 dataset, the original \meminf attack yielded metrics of 0.529 (Accuracy), 0.490 (F1 Score), 0.538 (AUC), and 0.001 (TPR @0.1\% FPR), respectively. When augmented with \adv, these values significantly increased to 0.659, 0.701, 0.674, and 0.004, respectively. 
Furthermore, for \tuple{\LIRA, \SD}, the \atoa{\adv}{\meminf} approach demonstrated consistent performance improvements across all three datasets.

%-------------------------------------------------------------------------------
\subsubsection{\atoa{\adv}{\propinf}}
%-------------------------------------------------------------------------------

\autoref{table:adv2propinf_dp} demonstrate the performance of \atoa{\adv}{\propinf} with adding DP.
From this table, we posit that the performance of models trained on CelebA can remain stable when DP is added during training. 
Specifically, the distribution of the $L_{2}$ distances between original samples and their corresponding adversarial examples assists in distinguishing models with different properties. 
In contrast, for the other two datasets, the model outputs inherently exhibit substantial noise, which in turn obscures the effect of DP and compromises the consistency of the evaluation.
Overall, while DP serves as a defense mechanism against \meminf, its influence on attack composition appears limited—unless it significantly degrades the model performance. 

\begin{table}[!t]
\centering
\caption{Performance of \atoa{\adv}{\propinf} with adding DP.}
\label{table:adv2propinf_dp}
\setlength{\tabcolsep}{4 pt}
\customTableFont
\begin{tabular}{@{}l l| c c c@{}}
\toprule
& {\bf Mode} & {\bf CelebA} & {\bf CIFAR10} & {\bf Places} \\
\midrule
\multirow{2}{*}{\bf $\epsilon = 10$} & Origin & 0.740 & 0.890 & 0.830 \\
& Composition & 0.790 & 0.610 & 0.670 \\
\midrule
\multirow{2}{*}{\bf $\epsilon = 20$} & Origin & 0.730 & 0.790 & 0.830 \\
& Composition & 0.770 & 0.540 & 0.790 \\
\midrule
\multirow{2}{*}{\bf $\epsilon = 50$} & Origin & 0.710 & 0.790 & 0.890 \\
& Composition & 0.740 & 0.670 & 0.780 \\
\bottomrule
\end{tabular}
\end{table}

%-------------------------------------------------------------------------------
\subsubsection{\atoa{\propinf}{\meminf}}
%-------------------------------------------------------------------------------

\begin{figure*}[!t]
\centering
\includegraphics[width=2.0\columnwidth]{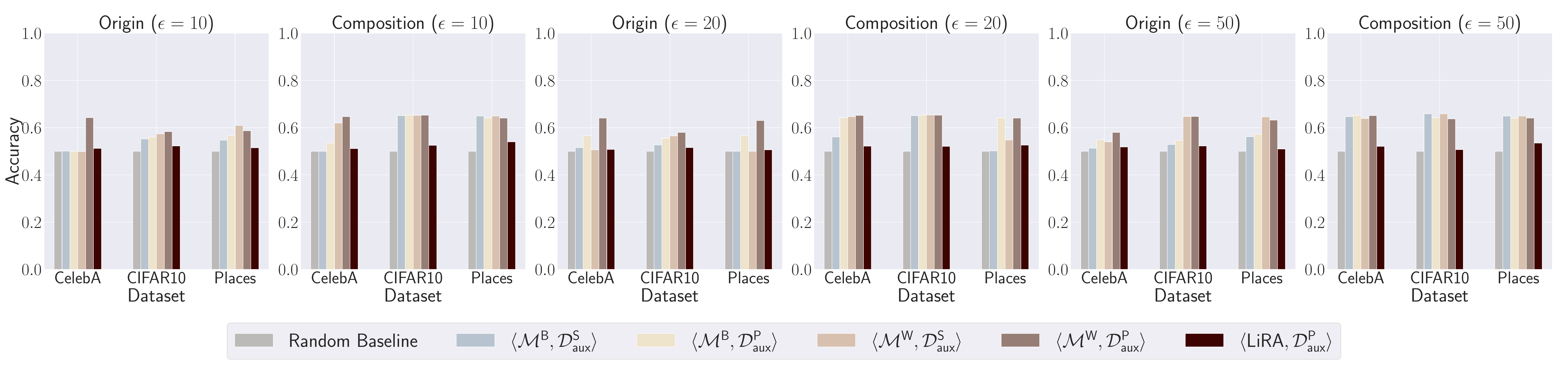}
\caption{Accuracy of \atoa{\propinf}{\meminf} under different threat models, datasets, and target model architectures with adding DP.}
\label{figure:propinf2meminf_acc_dp}
\end{figure*}

\autoref{figure:propinf2meminf_acc_dp}, \autoref{figure:propinf2meminf_f1_dp}, \autoref{figure:propinf2meminf_auc_dp}, and \autoref{table:tprpropinf2meminf_dp} demonstrate the results of \atoa{\propinf}{\meminf} after adding DP during the training phase.
Compared to the original \meminf, the \meminf calibrated with \propinf shows significant improvement, such as \tuple{\BM, \SD} and \tuple{\BM, \PD}, despite the addition of DP.
Therefore, we conclude that although adding DP can defend against \meminf, it is still possible to enhance \meminf through \propinf.

\begin{table*}[!t]
\centering
\caption{Performance of \atoatoa{\adv}{\propinf}{\attrinf}.
Here, we only show the empirical setting.}
\label{table:adv2propinf2attrinf}
\setlength{\tabcolsep}{4 pt}
\customTableFont
\begin{tabular}{@{}l l | c c c c c c@{}}
\toprule
& & \multicolumn{2}{c}{\bf CelebA} & \multicolumn{2}{c}{\bf CIFAR10} & \multicolumn{2}{c}{\bf Places} \\
{\bf Model} & {\bf Mode} & {\bf F1 Score} & {\bf Accuracy} & {\bf F1 Score} & {\bf Accuracy} & {\bf F1 Score} & {\bf Accuracy} \\
\midrule
\multirow{2}{*}{\bf Densenet121} & Origin & 0.771 & 0.712 & 0.916 & 0.911 & 0.667 & 0.500 \\
& Empirical  & 0.795 & 0.792 & 0.937 & 0.925 & 0.922 & 0.918 \\
\multirow{2}{*}{\bf ResNet18} & Origin & 0.667 & 0.500 & 0.667 & 0.500 & 0.667 & 0.500 \\
& Empirical & 0.790 & 0.782 & 0.901 & 0.901 & 0.918 & 0.896 \\
\multirow{2}{*}{\bf VGG19} & Origin & 0.667 & 0.500 & 0.667 & 0.500 & 0.667 & 0.500 \\
& Empirical & 0.764 & 0.761 & 0.938 & 0.938 & 0.923 & 0.919 \\

\bottomrule
\end{tabular}
\end{table*}

\mypara{Takeaways}
Despite DP being a very effective defense mechanism against \meminf, it can still be bypassed through auxiliary attacks that enhance its effectiveness. 
Furthermore, DP remains ineffective against other types of attacks, and when these other attacks are used as auxiliary methods.

%-------------------------------------------------------------------------------
\subsection{Chain of Composition} 
\label{section:chaincompose}
%-------------------------------------------------------------------------------

To facilitate a more in-depth and insightful discussion, we propose the concept of a chain of composition, namely \ensuremath{\mathsf{A_1}2\mathsf{A_2}2\mathsf{A_3}}. 
This approach leverages the interplay between different attacks to enhance the overall effectiveness of the attack strategy. 
By systematically composing multiple attacks, we aim to amplify their impacts on the target models.

%-------------------------------------------------------------------------------
\subsubsection{\atoatoa{\adv}{\propinf}{\attrinf}}
%-------------------------------------------------------------------------------

We only discuss the empirical settings here, since the theoretical setting is the same as \atoa{\propinf}{\attrinf}.
\autoref{table:adv2propinf2attrinf} demonstrates the results of \atoatoa{\adv}{\propinf}{\attrinf}.
We find that \atoatoa{\adv}{\propinf}{\attrinf} outperforms standalone \attrinf and is similar to \atoa{\propinf}{\attrinf} (see \autoref{section:preeva}). 
For example, the attack accuracy of standalone \attrinf on ResNet18 with CIFAR10 is 0.500, while \atoa{\propinf}{\attrinf} achieves 0.894, and \atoatoa{\adv}{\propinf}{\attrinf} achieves 0.901. 
\atoatoa{\adv}{\propinf}{\attrinf} performs similarly to \atoa{\propinf}{\attrinf} because the confidence of the predictions of \propinf with \adv enhancement is similar to that of standalone \propinf models. 
This similarity in confidence, exploited by \attrinf, results in comparable performance between \atoatoa{\adv}{\propinf}{\attrinf} and \atoa{\propinf}{\attrinf}. 
However, this does not contradict our claim that \atoa{\adv}{\propinf} outperforms standalone \propinf, as we measure the overall attack accuracy instead of the confidence.

\begin{figure*}[!t]
\centering
\includegraphics[width=2.0\columnwidth]{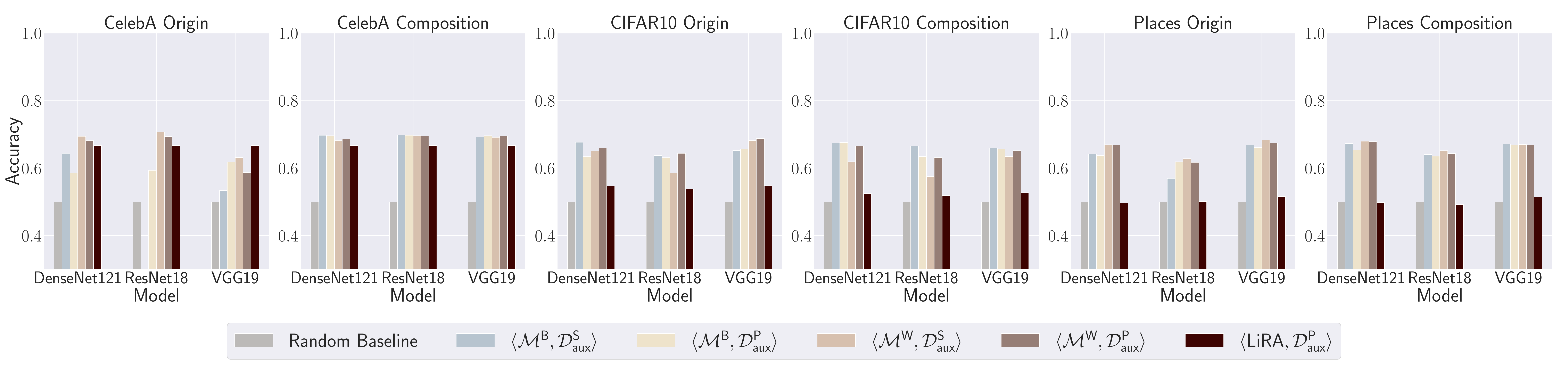}
\caption{Accuracy of \atoatoa{\adv}{\propinf}{\meminf} under different threat models, datasets, and target model architectures.}
\label{figure:adv2propinf2meminf_acc}
\end{figure*} 

%-------------------------------------------------------------------------------
\subsubsection{\atoatoa{\adv}{\propinf}{\meminf}}
%-------------------------------------------------------------------------------

\autoref{figure:adv2propinf2meminf_acc}, \autoref{figure:adv2propinf2meminf_f1}, and \autoref{figure:adv2propinf2meminf_auc} show the results of \atoatoa{\adv}{\propinf}{\meminf}.
\atoatoa{\adv}{\propinf}{\meminf} outperforms standalone \meminf and is similar to \atoa{\propinf}{\meminf} (see \autoref{section:evaleval}). 
For example, the accuracy of standalone \meminf on VGG19 with Places20 is 0.630, \atoa{\propinf}{\meminf} achieves 0.647, and \atoatoa{\adv}{\propinf}{\meminf} achieves 0.652. 
The reason that \atoatoa{\adv}{\propinf}{\meminf} performs similarly to \atoa{\propinf}{\meminf} is the same as discussed above: the confidence of the \atoa{\adv}{\propinf} model is similar to that of the \propinf model, which \meminf exploits.

\mypara{Takeaways}
The chain of composition is an interesting phenomenon that shows significant improvement compared to the original method. 
We will explore the depth and insights of our paper.

%-------------------------------------------------------------------------------
\section{Related Work} 
\label{section:relatedwork}
%-------------------------------------------------------------------------------

\mypara{Security and Privacy Attacks Against ML models}
ML models are vulnerable to various security and privacy attacks. 
Specifically, we focus on four representative attacks at the inference phase of the target ML model to better study the interactions of different attacks at broad levels. 
Adversarial examples are the most popular security attacks against ML models. 
They can mislead the prediction of the target ML model by adding imperceptible perturbations to the input data sample at the inference phase. 
The early methods~\cite{SZSBEGF14, GSS15} for finding adversarial examples assume that the adversary has full access to the target ML model (i.e., the white-box setting) and they rely on some optimization strategies to search for the results, while the latter variants~\cite{ACFH20, LXZYL20} have been developed to cope with various more challenging scenarios (e.g., the black-box setting). 
In this work, we implement the white-box attack method $\mathsf{PGD}$~\cite{MMSTV18} and the black-box attack method $\mathsf{Square}$~\cite{ACFH20} to study the effectiveness of adversarial examples in assisting other attacks. 

Membership inference is a popular privacy attack to determine whether a data sample exists in the training dataset of the target ML model. 
Shokri et al.~\cite{SSSS17} propose the first membership inference attack against black-box ML models. 
They train multiple shadow models on the shadow dataset to mimic the behavior of the target model and then train an attack model on the prediction posteriors of these shadow models to predict the membership of the input data. 
Salem et al.~\cite{SZHBFB19} relax the key assumptions of Shokri et al.~\cite{SSSS17} and introduce the model- and data-independent method to effectively conduct membership inference on black-box models. 
Since then, membership inference has been adapted to different settings (e.g., white-box~\cite{NSH19} and label-only~\cite{LZ21}), domains (e.g., computer vision~\cite{LZ21, LLHYBZ22}, recommender system~\cite{ZRWRCHZ21}, and unlearning system~\cite{CZWBHZ21}), and models (e.g., generative model~\cite{CYZF20} and multi-exit model~\cite{ LLHYBZ22}). 
We adapt the methods proposed by Salem et al.~\cite{SZHBFB19}, and Nasr et al.~\cite{NSH19} to implement black-box and white-box membership inference in our work. 

Attribute inference is another representative privacy attack that tries to learn extra information unrelated to the target task of the ML model. 
Melis et al.~\cite{MSCS19} introduced the first sample-level attribute inference attack targeting federated ML systems. 
Song and Shmatikov~\cite{SS20} demonstrated that the potential risks of attribute inference stem from the intrinsic overlearning characteristics of ML models. 
A common strategy for conducting attribute inference is to utilize an auxiliary dataset to build the connection between the model embedding and the target attribute. 

Property inference is a privacy attack aiming to infer the general property (e.g., data distribution) of the training dataset. 
The inferred property is usually unrelated to the main task of the target ML model. 
Prior work usually relies on the gradients~\cite{MSCS19} or the model parameters~\cite{GWYGB18} on the auxiliary dataset built with abundant shadow models to conduct property inference on small and simple target model architectures, which is computationally prohibitive for large and complex target models. 
In this work, we adapt the previous attack methods by concatenating the query posteriors to serve as the input features for the final attack model. 

\mypara{Interactions among Attacks}
Several studies have revealed relationships between different types of attacks.
Li et al.~\cite{LZ21} found a positive correlation between a sample's membership status and its robustness to adversarial noise.
They leveraged the differing adversarial noise magnitudes of members and non-members to mount a membership inference attack.
However, our work differs significantly from theirs. 
We integrate one attack into another at different phases, using information from one attack to enhance or amplify another. 
In contrast, Li et al. rely on adversarial example information as the only signal for membership inference, without incorporating its original signal.
Recently, Wen et al.~\cite{WMHGGC24} proposed a method to strengthen membership inference through training-phase data poisoning attacks. However, data poisoning is a training-time attack, while membership inference occurs during the inference phase. 
We emphasize that although it is possible for an attacker to launch attacks during both the training and inference phases, this assumption is overly strong.
As the first to systematically study the interactions between different attacks, we start only with the inference-time attack, as this is the most realistic scenario.
We also investigate the chain of composition, through \atoatoa{\adv}{\propinf}{\attrinf} and \atoatoa{\adv}{\propinf}{\meminf}. 
We find that the chain of composition can effectively provide a form of attack enhancement.

The prior work most closely related to ours is by Chen et al.~\cite{ZCSZ22}, who have found that property inference could amplify the performance of membership inference on GANs. However, their study focuses solely on GANs and proposes only one case study of attack composition.
Furthermore, it lacks a high-level awareness of the intentional interactions and does not provide a systematic study of the intentional interactions among a more diverse set of attacks.
Nonetheless, we acknowledge that their work provides valuable insights and inspires us to conduct this study.

%------------------------------------------------------------------------------

%-------------------------------------------------------------------------------
\section{Discussion} 
\label{section:discuss}
%------------------------------------------------------------------------------

We now discuss in more detail why we focus on the four inference-time attacks in the image domain.
There are some attacks during the training phase, such as enhancing membership inference through backdoor attacks~\cite{WMHGGC24} or poisoning attacks~\cite{CSSWZ22}. 
These situations are more complex, especially in real-world scenarios, requiring adversaries to make many strong assumptions, such as interfering with the training process or owning the training dataset. 
In addition, we currently only focus on image datasets because the types of attacks and their implementations are more detailed and comprehensive in image datasets.
We also do not consider model stealing attacks, as they primarily, to some extent, convert black-box models to white-box models, which indeed can amplify the success rate of many attacks. 
Since we aim to explore the impact of attack compositions during the attack's different phases, we emphasize that we do not change the overall attack process and the main attack approach. 
We emphasize that, currently, no single defense can protect against all ML model attacks, and effective defenses against property inference or attribute inference are lacking.
We aim to provide new insights and techniques for enhancing model security through these attack compositions.
We leave the in-depth exploration of more effective defense mechanisms against our attack compositions as future work.

%-------------------------------------------------------------------------------
\section{Conclusion} 
\label{section:conclusion}
%------------------------------------------------------------------------------

In this paper, we take the first step in exploring the intentional interaction between different types of attacks. 
Specifically, we focus on four extensively studied inference-time attacks: adversarial examples, attribute inference, membership inference, and property inference. 
To facilitate the study of their interactions, we establish a taxonomy based on three levels of the attack pipeline: preparation, execution, and evaluation, and propose four different attack compositions: \atoa{\propinf}{\attrinf}, \atoa{\adv}{\meminf}, \atoa{\adv}{\propinf}, and \atoa{\propinf}{\meminf}. 
Extensive experiments across three model architectures and two benchmark datasets demonstrate the superior performance of the proposed attack compositions.

Additionally, we introduce a reusable modular framework named \systemname to integrate our attack compositions. 
In this framework, we build four distinct modules to systematically examine the attack compositions. 
We believe that \systemname will serve as a benchmark tool to facilitate future research on attack compositions, enabling the seamless integration of new attacks, datasets, and models to further explore ML model vulnerabilities.

We explore DP as a defense mechanism to test our different attack compositions. 
We find that although DP can serve as a defense against \meminf, through our composition methods, we can still enhance the effectiveness of \meminf. 
Additionally, for other types of attacks, DP does not function as an effective defense mechanism.

Overall, we find that many current attacks can be amplified by applying another type of attack, thereby increasing their effectiveness. This finding offers a new perspective for subsequent attacks. 
In future work, we plan to incorporate more related attacks to further explore the vulnerabilities of ML models. 
With \systemname, we appeal to the community to consider such real-world scenarios and actively contribute to the development of more robust and secure AI systems.

%-------------------------------------------------------------------------------
\bibliographystyle{plain}
\bibliography{normal_generated_py3}
%-------------------------------------------------------------------------------

%-------------------------------------------------------------------------------
\appendix
% \section{Additional Experimental Results}
\label{section:appendix}
%-------------------------------------------------------------------------------

% In this appendix, we report plots for additional experiments as mentioned throughout the paper.

\begin{table*}[h]
\smallskip
\centering
\caption{TPR @0.1\% FPR of \atoa{\adv}{\meminf}.}
\label{table:tpradv2meminf}
\setlength{\tabcolsep}{3 pt}
\customTableFont
\begin{tabular}{@{}l l | c c c c c c@{}}
\toprule
& & \multicolumn{2}{c}{\bf CelebA} & \multicolumn{2}{c}{\bf CIFAR10} & \multicolumn{2}{c}{\bf Places} \\
{\bf Model} & {\bf Mode} & {\bf Origin} & {\bf Composition} & {\bf Origin} & {\bf Composition} & {\bf Origin} & {\bf Composition} \\
\midrule
\multirow{5}{*}{\bf DenseNet121} & $\tuple{\BM, \SD}$ & 0.000 & 0.007 & 0.009 & 0.011 & 0.002 & 0.003 \\
& $\tuple{\BM, \PD}$ & 0.002 & 0.006 & 0.002 & 0.217 & 0.002 & 0.003 \\
& $\tuple{\WM, \SD}$ & 0.004 & 0.008 & 0.016 & 0.887 & 0.001 & 0.500 \\
& $\tuple{\WM, \PD}$ & 0.004 & 0.010 & 0.011 & 0.875 & 0.002 & 0.486 \\
& $\tuple{\LIRA, \SD}$ & 0.054 & 0.079 & 0.105 & 0.203 & 0.061 & 0.182 \\
\midrule
\multirow{5}{*}{\bf ResNet18} & $\tuple{\BM, \SD}$ & 0.001 & 0.003 & 0.004 & 0.006 & 0.002 & 0.004 \\
& $\tuple{\BM, \PD}$ & 0.003 & 0.009 & 0.004 & 0.073 & 0.002 & 0.003 \\
& $\tuple{\WM, \SD}$ & 0.002 & 0.007 & 0.003 & 0.879 & 0.001 & 0.501 \\
& $\tuple{\WM, \PD}$ & 0.004 & 0.008 & 0.009 & 0.868 & 0.001 & 0.490 \\
& $\tuple{\LIRA, \SD}$ & 0.078 & 0.283 & 0.120 & 0.144 & 0.013 & 0.088 \\
\midrule
\multirow{5}{*}{\bf VGG19} & $\tuple{\BM, \SD}$ & 0.001 & 0.006 & 0.002 & 0.074 & 0.002 & 0.004 \\
& $\tuple{\BM, \PD}$ & 0.001 & 0.011 & 0.003 & 0.239 & 0.002 & 0.009 \\
& $\tuple{\WM, \SD}$ & 0.001 & 0.008 & 0.016 & 0.902 & 0.001 & 0.500 \\
& $\tuple{\WM, \PD}$ & 0.001 & 0.009 & 0.002 & 0.899 & 0.001 & 0.494 \\
& $\tuple{\LIRA, \PD}$ & 0.007 & 0.034 & 0.180 & 0.278 & 0.026 & 0.056 \\
\bottomrule
\end{tabular}
\end{table*}

\begin{table*}[h]
\centering
\caption{TPR @0.1\% FPR of \atoa{\propinf}{\meminf}.}
\label{table:tprpropinf2meminf}
\setlength{\tabcolsep}{4 pt}
\customTableFont
\begin{tabular}{@{}l l | c c c c c c@{}}
\toprule
& & \multicolumn{2}{c}{\bf CelebA} & \multicolumn{2}{c}{\bf CIFAR10} & \multicolumn{2}{c}{\bf Places} \\
{\bf Model} & {\bf Mode}& {\bf Origin} & {\bf Composition} & {\bf Origin} & {\bf Composition} & {\bf Origin} & {\bf Composition} \\
\midrule
\multirow{5}{*}{\bf DenseNet121} & $\tuple{\BM, \SD}$ & 0.002 & 0.007 & 0.003 & 0.002 & 0.003 & 0.003 \\
& $\tuple{\BM, \PD}$ & 0.001 & 0.007 & 0.000 & 0.001 & 0.003 & 0.003 \\
& $\tuple{\WM, \SD}$ & 0.002 & 0.009 & 0.000 & 0.001 & 0.003 & 0.002 \\
& $\tuple{\WM, \PD}$ & 0.003 & 0.009 & 0.000 & 0.000 & 0.002 & 0.003 \\
& $\tuple{\LIRA, \SD}$ & 0.003 & 0.003 & 0.002 & 0.002 & 0.001 & 0.003 \\
\midrule
\multirow{5}{*}{\bf ResNet18} & $\tuple{\BM, \SD}$ & 0.000 & 0.004 & 0.000 & 0.002 & 0.000 & 0.002 \\
& $\tuple{\BM, \PD}$ & 0.001 & 0.006 & 0.000 & 0.000 & 0.001 & 0.001 \\
& $\tuple{\WM, \SD}$ & 0.004 & 0.007 & 0.002 & 0.001 & 0.002 & 0.002 \\
& $\tuple{\WM, \PD}$ & 0.005 & 0.009 & 0.001 & 0.004 & 0.002 & 0.004 \\
& $\tuple{\LIRA, \SD}$ & 0.003 & 0.004 & 0.002 & 0.004 & 0.001 & 0.001 \\
\midrule
\multirow{5}{*}{\bf VGG19} & $\tuple{\BM, \SD}$ & 0.001 & 0.010 & 0.001 & 0.001 & 0.001 & 0.004 \\
& $\tuple{\BM, \PD}$ & 0.001 & 0.014 & 0.000 & 0.003 & 0.002 & 0.001 \\
& $\tuple{\WM, \SD}$ & 0.001 & 0.013 & 0.001 & 0.002 & 0.001 & 0.001 \\
& $\tuple{\WM, \PD}$ & 0.001 & 0.015 & 0.005 & 0.000 & 0.004 & 0.001 \\
& $\tuple{\LIRA, \SD}$ & 0.001 & 0.001 & 0.001 & 0.001 & 0.000 & 0.000 \\
\bottomrule
\end{tabular}
\end{table*}

\begin{table*}[h]
\centering
\caption{TPR @0.1\% FPR of \atoa{\propinf}{\meminf} with adding DP.}
\label{table:tprpropinf2meminf_dp}
\setlength{\tabcolsep}{4 pt}
\customTableFont
\begin{tabular}{@{}l l | c c c c c c@{}}
\toprule
& & \multicolumn{2}{c}{\bf CelebA} & \multicolumn{2}{c}{\bf CIFAR10} & \multicolumn{2}{c}{\bf Places} \\
& {\bf Mode} & {\bf Origin} & {\bf Composition} & {\bf Origin} & {\bf Composition} & {\bf Origin} & {\bf Composition} \\
\midrule
\multirow{5}{*}{\bf $\epsilon = 10$} & $\tuple{\BM, \SD}$ & 0.002 & 0.004 & 0.001 & 0.001 & 0.001 & 0.004 \\
& $\tuple{\BM, \PD}$ & 0.002 & 0.002 & 0.004 & 0.000 & 0.001 & 0.005 \\
& $\tuple{\WM, \SD}$ & 0.000 & 0.003 & 0.005 & 0.002 & 0.001 & 0.000 \\
& $\tuple{\WM, \PD}$ & 0.001 & 0.001 & 0.000 & 0.001 & 0.002 & 0.003 \\
& $\tuple{\LIRA, \SD}$ & 0.000 & 0.000 & 0.000 & 0.000 & 0.002 & 0.002 \\
\midrule
\multirow{5}{*}{\bf $\epsilon = 20$} & $\tuple{\BM, \SD}$ & 0.001 & 0.005 & 0.004 & 0.012 & 0.001 & 0.002 \\
& $\tuple{\BM, \PD}$ & 0.002 & 0.001 & 0.002 & 0.007 & 0.001 & 0.005 \\
& $\tuple{\WM, \SD}$ & 0.000 & 0.005 & 0.004 & 0.008 & 0.005 & 0.005 \\
& $\tuple{\WM, \PD}$ & 0.000 & 0.005 & 0.016 & 0.010 & 0.000 & 0.009 \\
& $\tuple{\LIRA, \SD}$ & 0.000 & 0.001 & 0.000 & 0.002 & 0.003 & 0.003 \\
\midrule
\multirow{5}{*}{\bf $\epsilon = 50$} & $\tuple{\BM, \SD}$ & 0.005 & 0.009 & 0.009 & 0.035 & 0.001 & 0.001 \\
& $\tuple{\BM, \PD}$ & 0.004 & 0.004 & 0.005 & 0.007 & 0.000 & 0.001 \\
& $\tuple{\WM, \SD}$ & 0.001 & 0.001 & 0.004 & 0.016 & 0.002 & 0.002 \\
& $\tuple{\WM, \PD}$ & 0.003 & 0.002 & 0.004 & 0.011 & 0.001 & 0.001 \\
& $\tuple{\LIRA, \SD}$ & 0.003 & 0.012 & 0.003 & 0.002 & 0.001 & 0.002 \\
\bottomrule
\end{tabular}
\end{table*}

\begin{figure*}[!t]
\centering
\includegraphics[width=2.0\columnwidth]{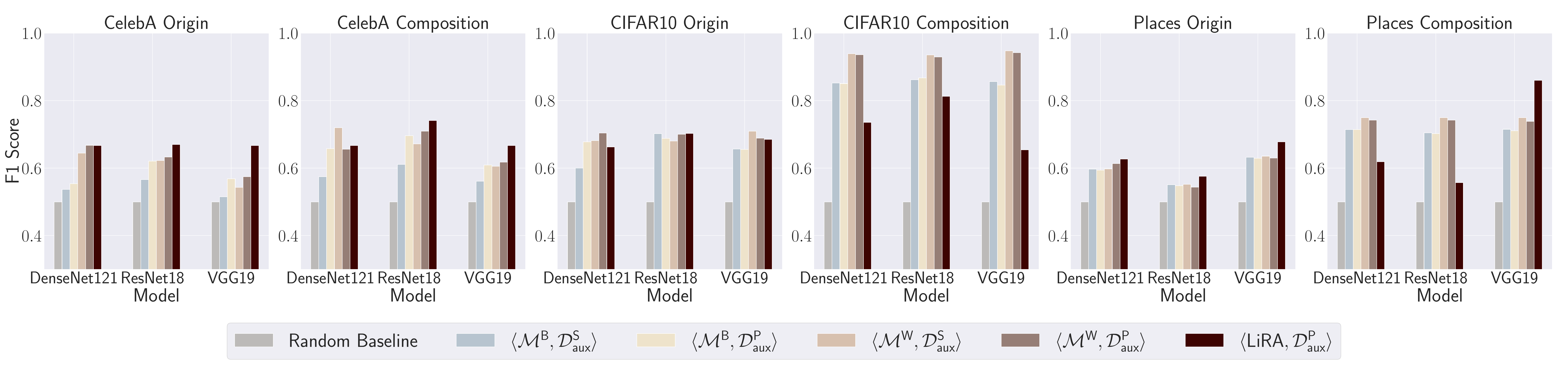}
\caption{F1 score of \atoa{\adv}{\meminf} under different threat models, datasets, and target model architectures.}
\label{figure:adv2meminf_f1}
\end{figure*} 

\begin{figure*}[!t]
\centering
\includegraphics[width=2.0\columnwidth]{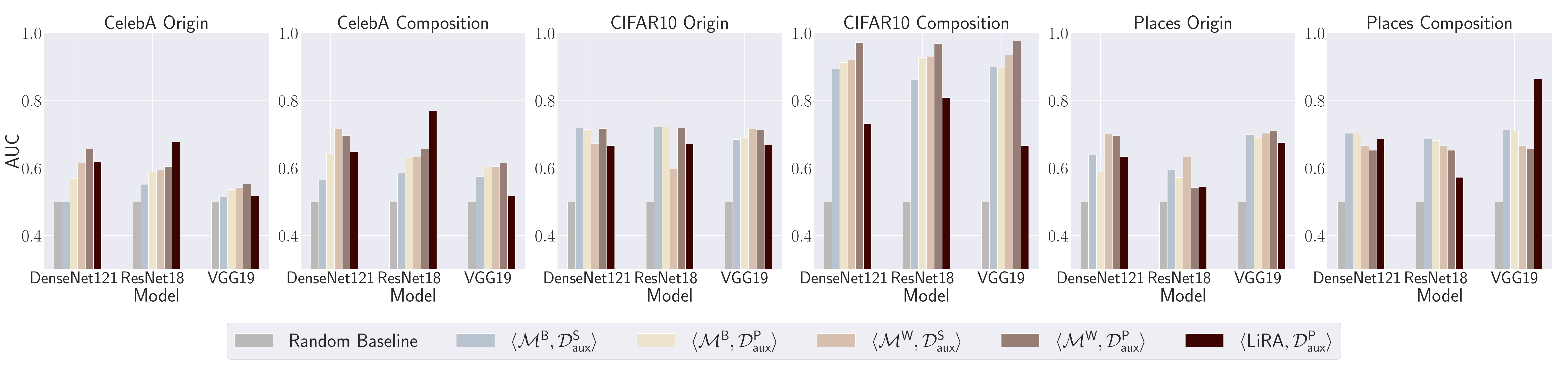}
\caption{AUC of \atoa{\adv}{\meminf} under different threat models, datasets, and target model architectures.}
\label{figure:adv2meminf_auc}
\end{figure*}

\begin{figure*}[!t]
\centering
\includegraphics[width=2.0\columnwidth]{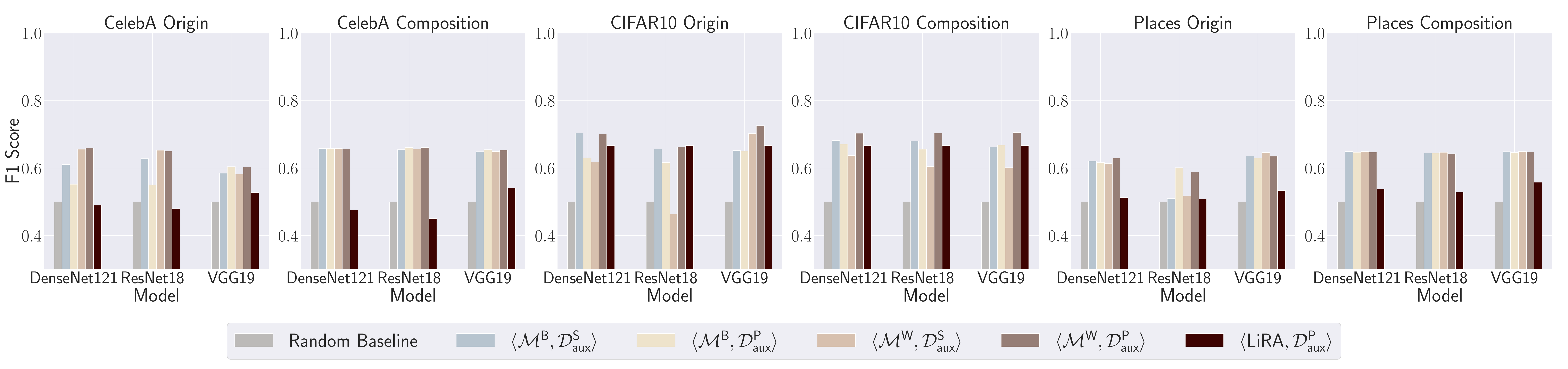}
\caption{F1 score of \atoa{\propinf}{\meminf} under different threat models, datasets, and target model architectures.}
\label{figure:propinf2meminf_f1}
\end{figure*} 

\begin{table*}[h]
\centering
\caption{TPR @0.1\% FPR of \atoa{\adv}{\meminf} with adding DP.}
\label{table:tpradv2meminf_dp}
\setlength{\tabcolsep}{4 pt}
\customTableFont
\begin{tabular}{@{}l l | c c c c c c@{}}
\toprule
& & \multicolumn{2}{c}{\bf CelebA} & \multicolumn{2}{c}{\bf CIFAR10} & \multicolumn{2}{c}{\bf Places} \\
& {\bf Mode}& {\bf Origin} & {\bf Composition} & {\bf Origin} & {\bf Composition} & {\bf Origin} & {\bf Composition} \\
\midrule
\multirow{5}{*}{\bf $\epsilon = 10$} & $\tuple{\BM, \SD}$ & 0.000 & 0.001 & 0.019 & 0.035 & 0.001 & 0.001 \\
& $\tuple{\BM, \PD}$ & 0.003 & 0.001 & 0.002 & 0.014 & 0.001 & 0.001 \\
& $\tuple{\WM, \SD}$ & 0.002 & 0.001 & 0.001 & 0.016 & 0.002 & 0.001 \\
& $\tuple{\WM, \PD}$ & 0.001 & 0.003 & 0.002 & 0.020 & 0.001 & 0.005 \\
& $\tuple{\LIRA, \SD}$ & 0.000 & 0.008 & 0.000 & 0.002 & 0.003 & 0.000 \\
\midrule
\multirow{5}{*}{\bf $\epsilon = 20$} & $\tuple{\BM, \SD}$ & 0.001 & 0.001 & 0.002 & 0.021 & 0.001 & 0.001 \\
& $\tuple{\BM, \PD}$ & 0.003 & 0.002 & 0.006 & 0.009 & 0.002 & 0.001 \\
& $\tuple{\WM, \SD}$ & 0.002 & 0.001 & 0.003 & 0.016 & 0.002 & 0.001 \\
& $\tuple{\WM, \PD}$ & 0.001 & 0.002 & 0.006 & 0.011 & 0.002 & 0.002 \\
& $\tuple{\LIRA, \SD}$ & 0.002 & 0.006 & 0.001 & 0.001 & 0.004 & 0.005 \\
\midrule
\multirow{5}{*}{\bf $\epsilon = 50$} & $\tuple{\BM, \SD}$ & 0.005 & 0.009 & 0.009 & 0.035 & 0.001 & 0.001 \\
& $\tuple{\BM, \PD}$ & 0.004 & 0.004 & 0.005 & 0.007 & 0.000 & 0.001 \\
& $\tuple{\WM, \SD}$ & 0.001 & 0.001 & 0.004 & 0.016 & 0.002 & 0.002 \\
& $\tuple{\WM, \PD}$ & 0.003 & 0.002 & 0.004 & 0.011 & 0.001 & 0.001 \\
& $\tuple{\LIRA, \SD}$ & 0.003 & 0.012 & 0.003 & 0.002 & 0.001 & 0.002 \\
\bottomrule
\end{tabular}
\end{table*}

\begin{figure*}[!t]
\centering
\includegraphics[width=2.0\columnwidth]{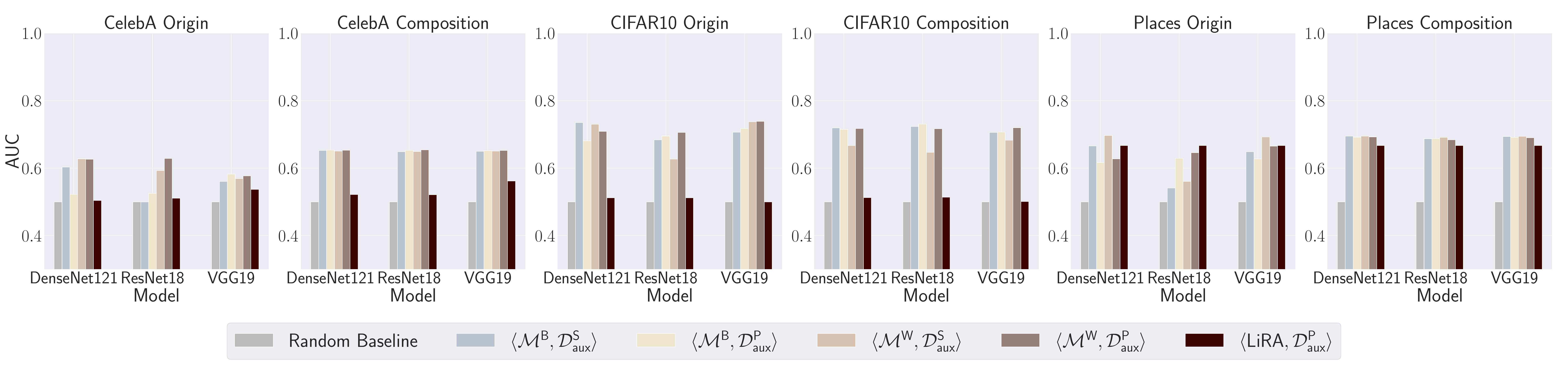}
\caption{AUC of \atoa{\propinf}{\meminf} under different threat models, datasets, and target model architectures.}
\label{figure:propinf2meminf_auc}
\end{figure*}

\begin{figure*}[!t]
\centering
\includegraphics[width=2.0\columnwidth]{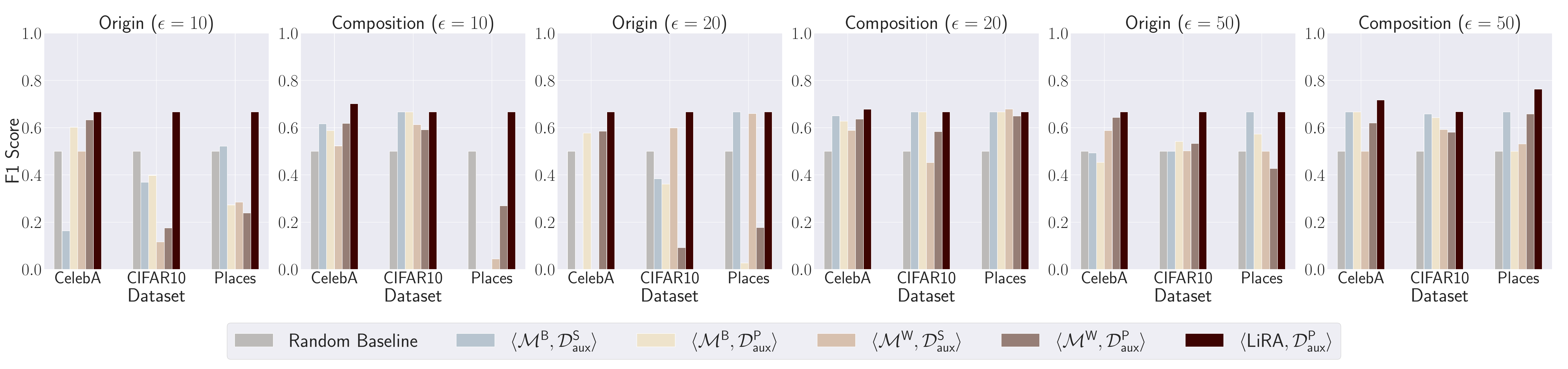}
\caption{F1 score of \atoa{\adv}{\meminf} under different threat models, datasets, and target model architectures with adding DP.}
\label{figure:adv2meminf_f1_dp}
\end{figure*} 

\begin{figure*}[!t]
\centering
\includegraphics[width=2.0\columnwidth]{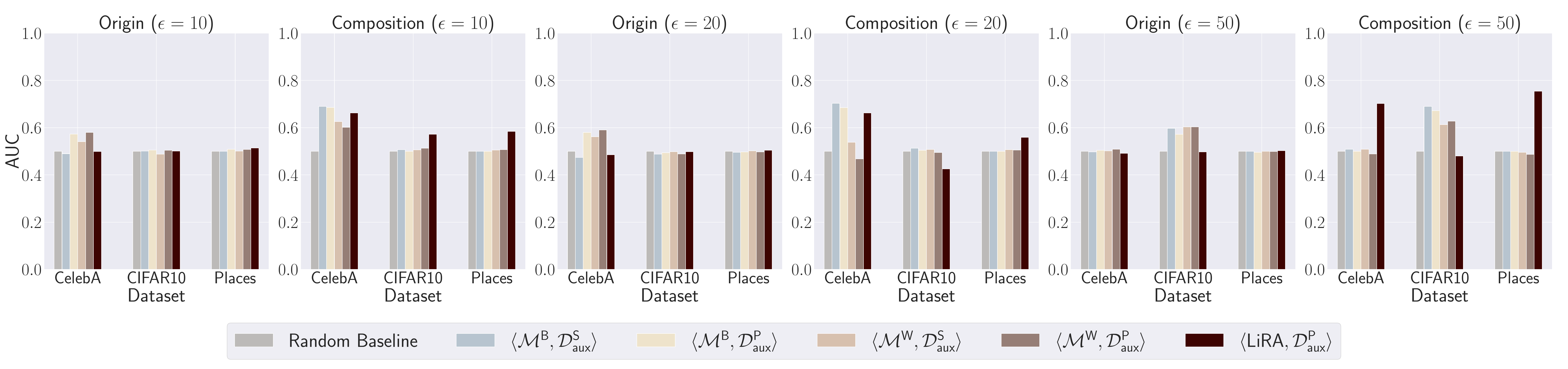}
\caption{AUC of \atoa{\adv}{\meminf} under different threat models, datasets, and target model architectures with adding DP.}
\label{figure:adv2meminf_auc_dp}
\end{figure*} 

\begin{figure*}[!t]
\centering
\includegraphics[width=2.0\columnwidth]{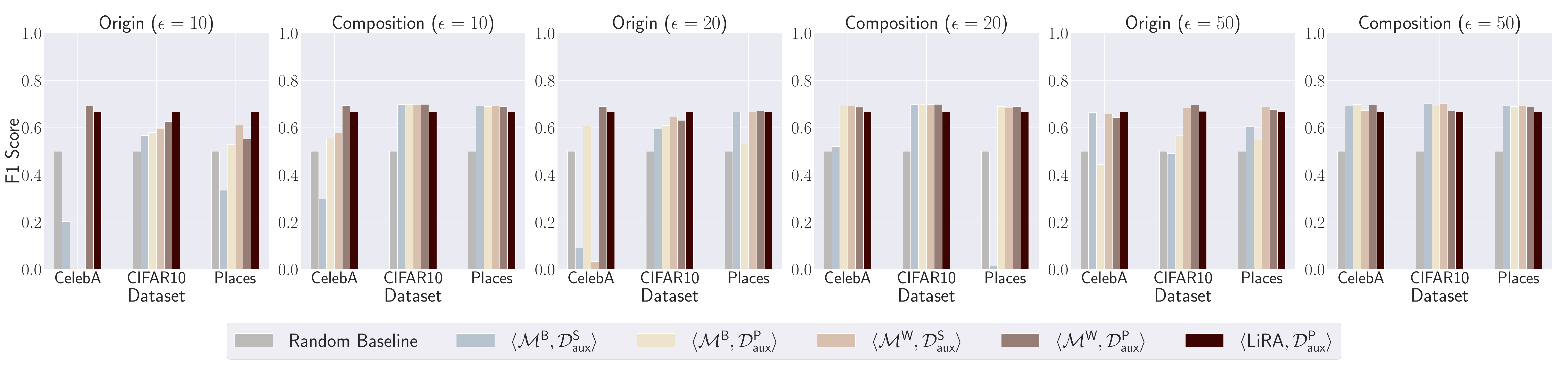}
\caption{F1 score of \atoa{\propinf}{\meminf} under different threat models, datasets, and target model architectures with adding DP.}
\label{figure:propinf2meminf_f1_dp}
\end{figure*} 

\begin{figure*}[!t]
\centering
\includegraphics[width=2.0\columnwidth]{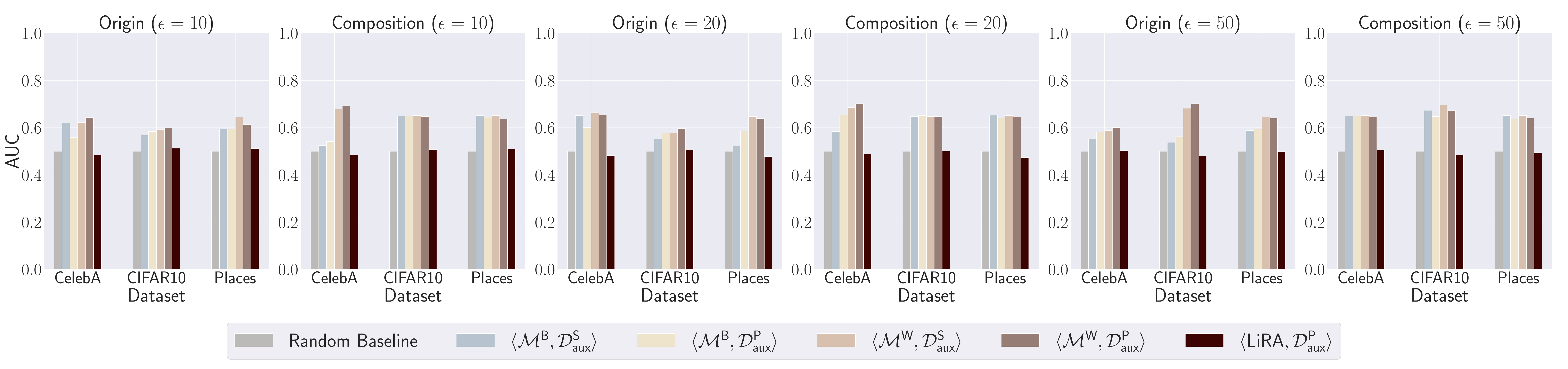}
\caption{AUC of \atoa{\propinf}{\meminf} under different threat models, datasets, and target model architectures with adding DP.}
\label{figure:propinf2meminf_auc_dp}
\end{figure*} 

\begin{figure*}[!t]
\centering
\includegraphics[width=2.0\columnwidth]{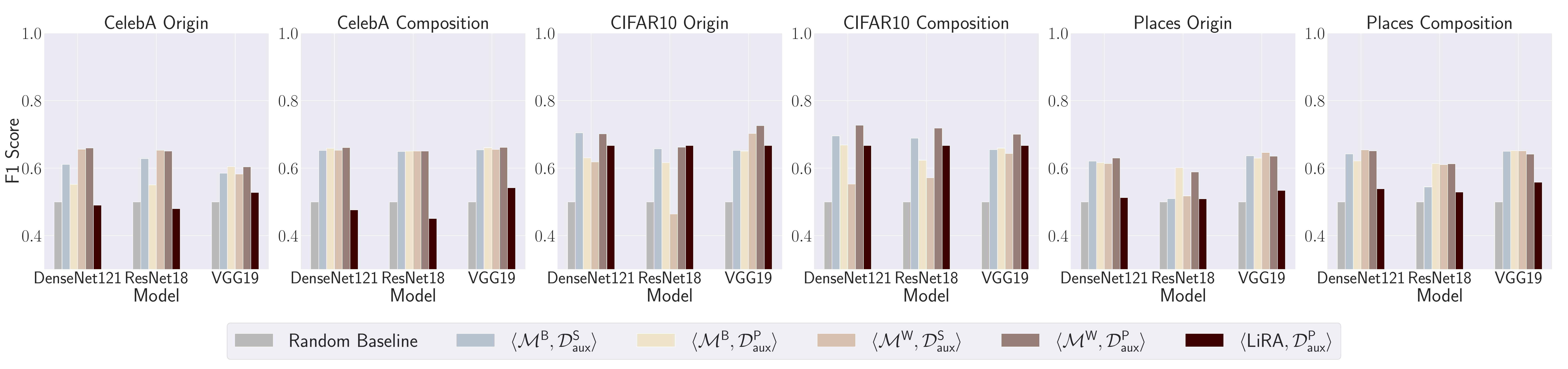}
\caption{F1 Score of \atoatoa{\adv}{\propinf}{\meminf} under different threat models, datasets, and target model architectures.}
\label{figure:adv2propinf2meminf_f1}
\end{figure*} 

\begin{figure*}[!t]
\centering
\includegraphics[width=2.0\columnwidth]{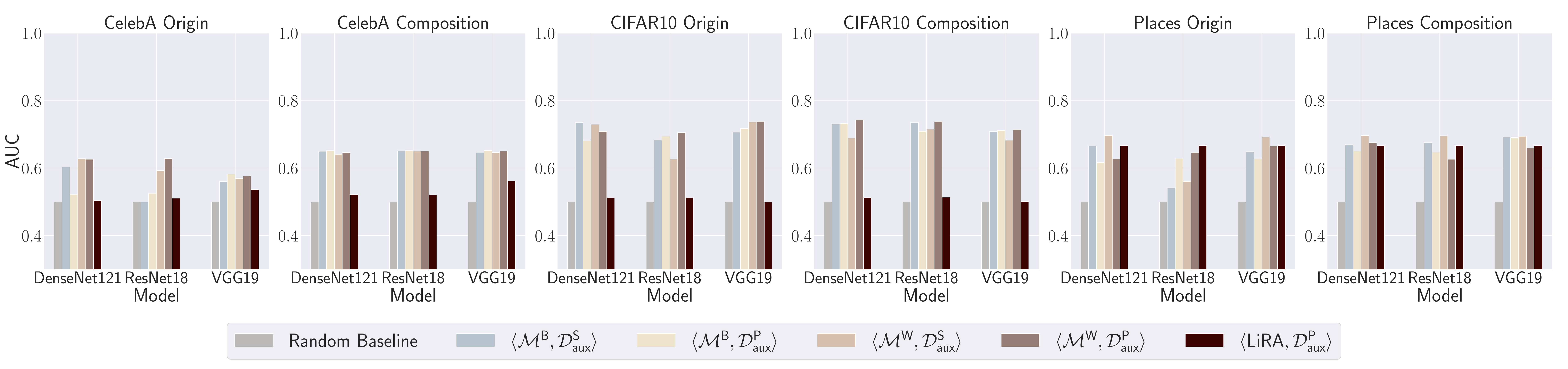}
\caption{AUC of \atoatoa{\adv}{\propinf}{\meminf} under different threat models, datasets, and target model architectures.}
\label{figure:adv2propinf2meminf_auc}
\end{figure*} 

%-------------------------------------------------------------------------------
\end{document}